\documentclass[review,12pt]{elsarticle}

\usepackage[english]{babel} 
\usepackage{csquotes} 
\usepackage{amsthm} 
\usepackage{xcolor,soul} 
\usepackage{siunitx} 
\usepackage{multirow} 
\usepackage{adjustbox} 
\usepackage{booktabs} 
\usepackage{hyphenat} 

\DeclareSIUnit{\atpercent}{at.\%}
\DeclareUnicodeCharacter{2212}{-}
\useshorthands{~}
\defineshorthand{~-}{\hyp{}}

\journal{Surface and Coatings Technology}

\begin{document}

\begin{frontmatter}

\title{Effect of Al content on the hardness and thermal stability study of AlTiN and AlTiBN coatings deposited by HiPIMS}


\author[1,2,3]{A. Mendez}
\author[2]{M.A. Monclus}
\author[3]{J.A. Santiago}
\author[3]{I. Fernandez-Martinez}
\author[4]{T.C. Rojas}
\author[2]{J. Garcia-Molleja}
\author[2]{M. Avella}
\author[5]{N. Dams}
\author[6,7]{M. Panizo-Laiz}
\author[2]{J.M. Molina-Aldareguia}

\address[1]{E.T.S. de Ingenieros de Caminos, Universidad Politecnica de Madrid, c/Prof. Aranguren s/n, 28040 Madrid, Spain}
\address[2]{IMDEA Materials Institute, c/Eric Kandel 2, 28906 Getafe, Spain}
\address[3]{Nano4Energy SL, c/ José Gutiérrez Abascal 2, 28006 Madrid, Spain}
\address[4]{Instituto de Ciencia de Materiales de Sevilla (CSIC-Univ. Sevilla, Avda. Américo Vespucio, 49, 41092 Sevilla, Spain}
\address[5]{PVT Plasma und Vakuum Technik GmbH, Rudolf-Diesel-Strasse 7, Bensheim 64625, Germany}
\address[6]{Departamento de Física Aplicada e Ingeniería de Materiales, E.T.S. de Ingenieros Industriales, Universidad Politécnica de Madrid, José Gutiérrez Abascal 2, 28006 Madrid, Spain}
\address[7]{Instituto de Fusión Nuclear ``Guillermo Velarde", c/ José Gutiérrez Abascal, 2, 28006 Madrid, Spain}

\begin{abstract}

The microstructure, mechanical properties and thermal stability of Al$_x$Ti$_{1-x}$N and Al$_{1−x}$Ti$_x$BN coatings grown by reactive high-power impulse magnetron sputtering (HiPIMS) have been analyzed as a function of $Al/(Al+Ti)$ ratio ($x$) between 0.5 and 0.8. The coatings were predominantly formed by a face-centered cubic Ti(Al)N crystalline phase, both with and without B, even for $x$ ratios as high as 0.6, which is higher than the ratio typically encountered for Al$_x$Ti$_{1-x}$N coatings deposited by reactive magnetron sputtering. B doping, in combination with the highly energetic deposition conditions offered by HiPIMS, results in the suppression of the columnar grain morphology typically encountered in Al$_x$Ti$_{1-x}$N coatings. On the contrary, the Al$_x$Ti$_{1-x}$BN coatings grown by HiPIMS present a dense nanocomposite type microstructure, formed by nanocrystalline Ti(Al)N domains and amorphous regions composed of Ti(Al)B$_2$ and BN. As a result, high-Al content ($x\approx0.6$) Al$_x$Ti$_{1-x}$BN coatings grown by HiPIMS offer higher hardness, elasticity and fracture toughness than Al$_x$Ti$_{1-x}$N coatings. Moreover, the thermal stability and the hot hardness are substantially enhanced, delaying the onset of formation of the detrimental hexagonal AlN phase from \SI{850}{\celsius} in the case of Al$_{0.6}$Ti$_{0.4}$N, to \SI{1000}{\celsius} in the case of Al$_{0.6}$Ti$_{0.4}$BN.

\end{abstract}



\begin{keyword}



coating \sep nitride \sep nanocomposite \sep high temperature \sep hardness \sep HiPIMS

\end{keyword}

\end{frontmatter}



\section{Introduction}\label{sec:Introduction}

Aluminum titanium nitride (Al$_{x}$Ti$_{1-x}$N) coatings are currently the most versatile coatings used in industry for many applications. They offer a superior wear, oxidation and corrosion resistance over conventional TiN coatings, which contributes to improve the durability of cutting tools \cite{kalss2006}. This superior behavior is typically found for high Al contents with $Al/(Al+Ti)$ ratios ($x$) up to 0.6, for which the coatings are composed by a single solid-solution Ti(Al)N cubic phase \cite{munz1986}.However, higher $x$ ratios typically lead to the appearance of the stable wurtzite-type hexagonal AlN phase, which is detrimental for the hardness of the coatings \cite{wahlstrom1993, zhou1999, kimura1999}. Adibi \textit{et al.} \cite{adibi1991} showed that single-phase cubic Al$_{x}$Ti$_{1-x}$N films can undergo a \textit{surface-initiated} spinodal decomposition, when deposited at high temperatures (over \SI{540}{\celsius}) resulting in the formation of TiN- and AlN-rich domains; a phenomenon that has been referred to as age-hardening. This phenomenon was also reported by Mayrhofer \textit{et al.} \cite{mayrhofer2003} when Al$_{x}$Ti$_{1-x}$N coatings were annealed over \SI{800}{\celsius}, marking the beginning of film decomposition. However, higher temperatures typically lead to loss of hardness due to the subsequent phase transformation to the stable AlN wurtzite structure.

The state of the art in the development of nitride coatings has recently focused on the alloying or doping of the well-established Al$_{x}$Ti$_{1-x}$N coating system to improve its high temperature performance even more. Various elements have been tested so far to fulfill this purpose yielding different degrees of performance, with Cr and Si standing out as the most successful ones. Cr has been shown to improve hardness and thermal resistance up to \SI{1000}{\celsius} \cite{yamamoto2003} and the machining tests results were also satisfactory \cite{foxrabinobich2009}. Si, on the other hand, has also proven to be successful in improving thermal resistance by triggering the formation of the so-called nanocomposite microstructure, composed of Ti(Al)N cubic phase nanograins surrounded by a tissue layer of amorphous SiN$_x$, as shown by Veprek \textit{et al.} \cite{veprek1995, veprek1999, veprek2004}.

More recently, boron has also been tested as an element to enhance the properties of TiN and AlTiN coatings. In 1989, Mitterer \textit{et al.} suggested that boron might also trigger microstructural refinement \cite{fan2015, wang2016}, which could make boron nitride-based materials very promising as tribological coatings \cite{mitterer1989}. Following this research line, Rebholz \textit{et al.} deposited AlTiBN films with high B contents (from 30 to \SI{52}{\atpercent}) with hardness values up to \SI{25}{\giga\pascal} \cite{rebholz2006}. Then, in 2007, they finally presented hard and superhard TiAlBN coatings deposited by electron-beam evaporation, showing outstanding results for boron contents of \SI{6}{\atpercent} in Al$_x$Ti$_{1-x}$BN coatings with low Al ratios ($x = 0.1$). They also estimated that the optimum boron content should be close to \SI{10}{\atpercent}  \cite{rebholz2007}. Kutschej \textit{et al.} also reported high hardness for TiAlBN coatings exhibiting a completely hexagonal matrix lattice \cite{kutschej2005}. Since then, other groups have also reported good hardness and oxidation resistance results results on TiAlBN-based coatings \cite{pfeiler2009, park2009, luong2017, chang2019}.

In all these studies, it was clear that the growth of superhard B-doped nitrides requires intense ion bombardment of the growing film \cite{veprek1999}. As a result, cathodic arc evaporation has become the industry-standard for the deposition of this type of coatings over magnetron sputtering owing to the high deposition rates that can be achieved and the high energy at which the particles reach the surface of the growing film. However, novel techniques like High-Power Impulse Magnetron Sputtering (HiPIMS) can close the gap with arc evaporation, by facilitating the generation of highly energetic target ions on top of the sputtered gas ions through the application of high-energy pulses in a short time and low duty cycle. The objective of this work was to assess the microstructure, mechanical properties and thermal stability of $Al_xTi_{1-x}(B)N$ coatings grown by HiPIMS. More specifically, the work aimed to address several open issues in the field. Firstly, to study the capability of HiPIMS to produce dense and compact Al$_x$Ti$_{1-x}$BN coatings with high Al contents. And second, to assess whether B doping results in coatings with superior mechanical properties and thermal stability, with respect to their Al$_x$Ti$_{1-x}$N counterparts. For that, the Al content was systematically varied in two series of Al$_x$Ti$_{1-x}$BN and Al$_x$Ti$_{1-x}$N coatings, grown using identical HiPIMS conditions. In each series. the Al content ($x$ ranging between 0.5 and 0.8) was adjusted by co-sputtering from a pre-alloyed target and a supporting target of either Al or Ti. The results show that B doping, in combination with the highly energetic deposition conditions offered by HiPIMS, increases considerably the hardness and fracture toughness of the coatings, and that the optimum Al content occurs for $x = 0.6$. Moreover, B doping enhances the thermal stability of the coatings. The results are explained based on the microstructure evolution with temperature as a function of Al and B content.

\section{Experimental details}\label{sec:Experimental}

\begin{figure}
		\centering
		\includegraphics[width=0.8\textwidth]{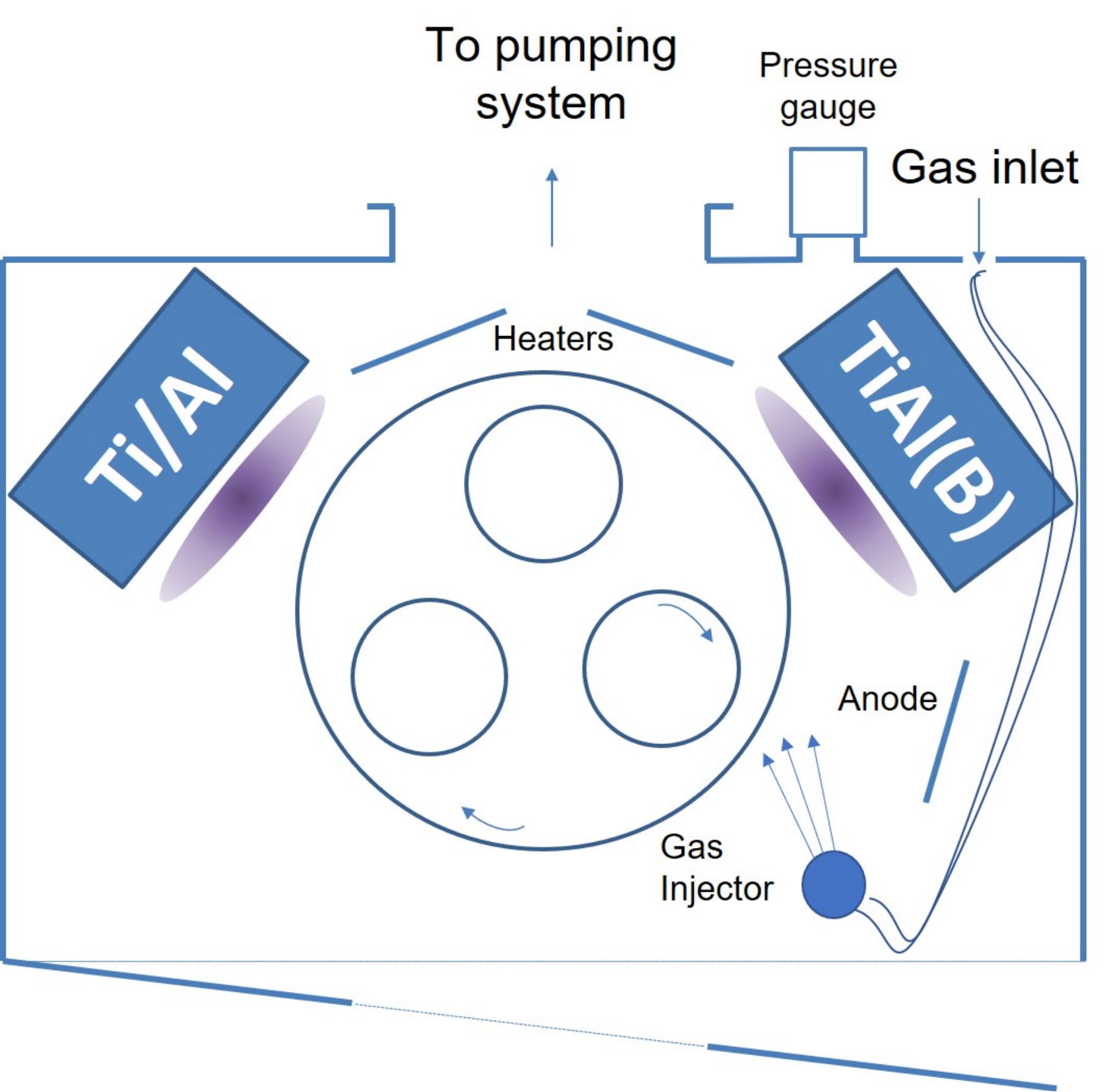}
		\caption{Schematics of the industrial-scale chamber used for sample deposition.}
	\label{fig:Chamber}
\end{figure}

The coatings were sputtered in an industrial scale vacuum chamber fitted with two rectangular magnetrons of size $400 \times 100$ \si{\square\milli\meter}, shutters, a two-fold rotating substrate holder with bias current, anode, heaters, mass flow controllers, and a pumping system (Fig. \ref{fig:Chamber}). Pre-alloyed Al$_{50}$Ti$_{50}$ and Al$_{55}$Ti$_{35}$B$_{10}$ powder-pressed ceramic targets were used for the deposition of the Al$_{x}$Ti$_{1-x}$N and Al$_{x}$Ti$_{1-x}$BN series, respectively. These targets were always placed on the right magnetron. The deposition was assisted by pure Al (99.99\%) or pure Ti (99.99\%) targets placed on the left magnetron to tailor the final $x$ ratio in each coating series.

Coatings were deposited on mirror electropolished (average roughness $R_a < 5$ \si{\nano\meter}) stainless steel (AISI304) substrates with a size of $40 \times 40$ \si{\square\milli\meter}, \SI{100}{\micro\meter} thick $36\times 6$ \si{\square\milli\meter} stainless steel (AISI420) substrates and on commercial (100) silicon substrates. Prior to deposition, the substrates were cleaned with soap and anti-grease product, then rinsed with water and lastly with iso-propyl alcohol (IPA), to finally dry them with filtered compressed air. Once inside the chamber, and after reaching a base pressure bellow \SI{e-2}{\pascal}, the substrates were argon etched using a DC-Pulsed substrate bias voltage with an average voltage of \SI{-450}{\volt} and a frequency of \SI{150}{\kilo\hertz}. After argon etching, a layer of Ti was deposited by HiPIMS. The pressure during deposition, \SI{0.7}{\pascal}; the N$_2/$Ar gas flow, 23/66 sccm; the deposition temperature, \SI{350}{\celsius}; the substrate bias of \SI{-100}{\volt}; and the substrate holder rotation speed, 5 rpm, were the same for all coatings.

In the case of the Al$_x$Ti$_{1-x}$N coating series, the Al$_{50}$Ti$_{50}$ target was operated at a power of \SI{3}{\kilo\watt} with a pulse duration of \SI{100}{\micro\second} and a frequency of \SI{750}{\hertz}. In order to adjust the Al content, different coatings were deposited with the supporting Al target operated at powers of 0.25, 0.5 and \SI{2}{\kilo\watt}, with a pulse duration of \SI{100}{\micro\second} and a frequency of \SI{750}{\hertz}; or with the supporting Ti target operated at \SI{0.5}{\kilo\watt}, with a pulse duration of \SI{150}{\micro\second} and a frequency of \SI{750}{\hertz}. A total deposition time of \SI{2.5}{\hour} was used for all coatings in this series.

In the case of the Al$_x$Ti$_{1-x}$BN coatings, the Al$_{55}$Ti$_{35}$B$_{10}$ target was operated at a power of \SI{3}{\kilo\watt} with a pulse duration of \SI{150}{\micro\second} and a frequency of \SI{550}{\hertz}. In order to adjust the Al content, different coatings were deposited with the supporting Ti target operated at powers of 0.5, 1.5 and \SI{2}{\kilo\watt}. The HiPIMS pulse parameters of the supporting target were kept constant in all cases, with a pulse duration of \SI{150}{\micro\second} and a frequency of \SI{750}{\hertz}. A total deposition time of \SI{2}{\hour} was used for all coatings in this series.

An additional AlTiBN coating was also deposited using DC magnetron sputtering (referred to as AlTiBN-DC), with the objective of producing a reference coating that could be used to assess the benefits of HiPIMS over conventional magnetron sputtering. In order to reproduce the setup from the best Al$_x$Ti$_{1-x}$BN coating obtained with HiPIMS, the same targets and sputtering conditions were used (\textit{i.e.}, Al$_{55}$Ti$_{35}$B$_{10}$ pre-alloyed target, operated at a power of \SI{3}{\kilo\watt} and the support Ti target, operated at a power of \SI{1}{\kilo\watt}).

Annealing tests were carried out under vacuum inside a Nabertherm High-Temperature Tube Furnace at 800, 850, 900, 1000 and \SI{1100}{\celsius}. Temperature was ramped up at a \SI{5}{\celsius\per\minute} rate up to the target setpoint. The samples were stabilized at the target temperature for \SI{30}{\minute} and finally the furnace was turned off and allowed to cool freely. The samples were characterized after each of the heat treatments and put back into the furnace for subsequent annealing at the next temperature level.

Compositional analysis was performed using Glow Discharge Optical Emission Spectrometry (GDOES) in a GD Profiler 2 from HORIBA. Determination of coating stress $\sigma$ was achieved by measuring the curvature of the substrate with a Sensofar Metrology S neox confocal microscope and applying Stoney's equation \cite{stoney1909}:

\begin{equation}
\sigma = \frac{E_sh_s^2}{6(1-\nu_s)h_fR}
\label{eq:Stoney}
\end{equation}

where $E_s$, $h_s$ and $\nu_s$ are the elastic modulus, thickness and Poisson's ratio from the substrate, $h_f$ is the coating thickness and $R$ is the curvature radius. For this purpose, thin \SI{100}{\micro\meter} stainless steel substrates (AISI420) were used, assuming a \SI{200}{\giga\pascal} elastic modulus and a 0.27 Poisson's ratio. The phases and textures present in the coatings were examined using a PANalytical Empyrean X~-ray diffractometer in the Bragg-Brentano configuration using CuK$\alpha$ radiation, operated at 45 \si{\kilo\volt} and 40 \si{\milli\ampere} with a Ni filter. As for the data treatment, the background was subtracted, each spectrum was normalized in relation to the highest peak belonging to the coating material and substrate peaks were matched to a steel diffraction pattern to evaluate peak shifts \cite{steelpattern}, with the use of the HighScore Plus software and the PDF-4+ 2020 ICDD database. The microstructure of the most relevant coatings was examined using a transmission electron microscope (TEM) (FEI Talos F200X) operated at \SI{200}{\kilo\volt}. Cross-sectional specimens for TEM analysis were prepared using a FEI Helios Nanolab 600i dual beam focused ion beam - field emission gun scanning electron microscope (FIB-FEGSEM). Electron energy loss spectroscopy (EELS) was performed using a Gatan Imaging Filter (GIF), model QUAMTUM, attached to a FEI Tecnai G2 F30 TEM working at \SI{300}{\kilo\volt}, and equipped with a high angle annular dark field (HAADF) detector from Fischione with a \SI{0.16}{\nano\meter} point resolution. EELS spectra were recorded in scanning TEM (STEM) mode using a probe with a size of less than \SI{1}{\nano\meter} and a spectrometer collection angle of \SI{12.4}{\milli\radian}. Under these conditions, the energy resolution of the coupled microscope/spectrometer system was \SI{\sim1}{\electronvolt}. After experimental acquisition, the data were processed using Gatan Digital Micrograph software for EELS spectra background subtraction and quantification. The mechanical properties were obtained by nanoindentation using a Hysitron TI950 triboindenter equipped with a Berkovich-shaped diamond tip, whose area function was calibrated using multiple indents on a standard fused silica sample. Penetration depths were kept below $10\%$ of the total film thickness in order to minimize substrate effects. At least 10 indents per sample, with maximum loads of \SI{12}{\milli\newton} were performed. For high temperature nanoindentation testing, a hot stage (Hysitron xSol) and a Berkovich diamond indenter fitted to a special long insulating shaft were used. In this configuration, the sample is placed between two resistive heating elements in order to eliminate temperature gradients across the sample thickness. Dry air and argon around the tip and sample surface were used to purge the testing area to prevent heated gases reaching the transducer and reduce possible oxidation. Once the sample reached the selected temperature (250, 500, 400, 600 and \SI{700}{\celsius}) and was stable at the target temperature to within \SI{\pm 0.1}{\celsius}, the tip was placed at a distance of \SI{100}{\micro\meter} from the sample surface for \SI{15}{\min}, to ensure passive heating of the tip before the start of the test and to minimize thermal drift. Coating hardness and modulus values were determined using the Oliver and Pharr method \cite{oliverpharr1992}. In order to calculate the elastic modulus of the coatings, a Poisson ratio of 0.25, similar to the TiN value, was assumed for all the coatings.

\section{Results}\label{sec:Results}

\subsection{Deposition rate, composition and coating stress}\label{subsec:Rate}

Table \ref{tab:Composition} summarizes the experimental deposition conditions used for each coating, the total thickness achieved and their chemical composition, as extracted from the GDOES measurements. The coatings were named in order of increasing $x$ ratio. In the case of the AlTiN series, deposition using only the pre-alloyed Al$_{50}$Ti$_{50}$ target led to an $x$ ratio of 0.54 (AlTiN-2). In this case, co-deposition with a supporting Ti target was used to reduce the $x$ ratio down to 0.51 for AlTiN-1, while the rest of the series was grown with the Al supporting target to reach an $x$ ratio of 0.78 (AlTiN-5). On the contrary, for the AlTiBN series, due to the high Al content of the pre-alloyed target (Al$_{55}$Ti$_{35}$B$_{10}$), the highest $x$ ratio was obtained for AlTiN-5 (0.69), without the use of any supporting target, while lower ratios, down to 0.56 (AlTiN-1), were reached with the contribution of the Ti supporting target. According to the GDOES measurements, all the nitride coatings were fairly stoichiometric or slightly over-stoichiometric, with $N/(Ti+Al)$ ratios slightly over 1. The B content obtained from the GDOES measurements in the AlTiBN series was not reliable because this technique is not optimized for very light elements, like B. However, the EELS measurements gave an estimate the B content, as will be discussed below.

The calculated coating stresses using Eq. \ref{eq:Stoney} were \SI{4.43}{\giga\pascal} for Al$_{0.59}$Ti$_{0.41}$N ($R$: \SI{120.4}{\milli\meter} and $h_f$: \SI{856.5}{\nano\meter}) and \SI{3.03}{\giga\pascal} for Al$_{0.62}$Ti$_{0.38}$BN ($R$: \SI{169.9}{\milli\meter} and $h_f$: \SI{885.7}{\nano\meter}): a $\approx32\%$ reduction in stress when adding B.

\begin{table}
	\centering
	\begin{adjustbox}{max width=\textwidth}
		\begin{tabular}{ l l l l l l l c c c }

			\toprule
			
			\multicolumn{1}{l}{\multirow{2}{*}{Coating}} & \multicolumn{1}{l}{\multirow{2}{*}
			{Targets}} & \multicolumn{1}{l}{\multirow{2}{*}{Power}} & \multicolumn{4}{c} 
			{Composition (\si{\atpercent})} & \multicolumn{1}{l}{\multirow{2}{*}{$x$}} & Thickness & Deposition Rate \\
			
			\cline{4-7}
			
			\multicolumn{1}{c}{} & \multicolumn{1}{c}{} & \multicolumn{1}{c}{} & \multicolumn{1}
			{c}{\multirow{1.3}{*}{Al}} & \multicolumn{1}{c}{\multirow{1.3}{*}{Ti}} & 
			\multicolumn{1}{c}{\multirow{1.3}{*}{B}} & \multicolumn{1}{c}{\multirow{1.3}{*}{N}}
			 & \multicolumn{1}{c}{} & \multicolumn{1}{c}{(\si{\micro\meter})} & \multicolumn{1}{c}{($\si{\micro\meter}/\si{\hour}$)}\\
			
			\midrule

			AlTiN-1 & Al$_{50}$Ti$_{50}$ + Ti & 3\si{\kilo\watt} and 0.5\si{\kilo\watt} & 24.40 & 23.80 & --- & 
			51.73 & 0.51 & 2.45 & 0.98 \\ 
			AlTiN-2 & Al$_{50}$Ti$_{50}$ & 3\si{\kilo\watt} & 25.79 & 21.85 & --- & 52.27 & 0.54 & 1.99 & 0.80 \\
			AlTiN-3 & Al$_{50}$Ti$_{50}$ + Al & 3\si{\kilo\watt} and 0.25\si{\kilo\watt} & 26.79 & 18.67 & --- & 
			54.55 & 0.59 & 1.59 & 0.64 \\
			AlTiN-4 & Al$_{50}$Ti$_{50}$ + Al & 3\si{\kilo\watt} and 0.5\si{\kilo\watt} & 29.36 & 14.98 & --- & 
			55.39 & 0.66 & 1.69 & 0.68 \\ 
			AlTiN-5 & Al$_{50}$Ti$_{50}$ + Al & 3\si{\kilo\watt} and 2.0\si{\kilo\watt} & 34.61 & 9.80 & ---  & 
			55.25 & 0.78 & 2.75 & 1.1 \\

			\addlinespace

			AlTiBN-1 & Al$_{55}$Ti$_{35}$B$_{10}$ + Ti & 3\si{\kilo\watt} and
			2.0\si{\kilo\watt} & 25.35 & 20.12 & 0.85 & 53.43 & 0.56 & 2.66 & 1.33 \\
			AlTiBN-2 & Al$_{55}$Ti$_{35}$B$_{10}$ + Ti & 3\si{\kilo\watt} and
			1.5\si{\kilo\watt} & 26.69 & 19.35 & 0.91 & 52.35 & 0.58 & 2.01 & 1.01 \\
			AlTiBN-3 & Al$_{55}$Ti$_{35}$B$_{10}$ + Ti & 3\si{\kilo\watt} and
			1.0\si{\kilo\watt} & 28.44 & 17.50 & 0.97 & 52.58 & 0.62 & 1.80 & 0.90 \\
			AlTiBN-4 & Al$_{55}$Ti$_{35}$B$_{10}$ + Ti & 3\si{\kilo\watt} and
			0.5\si{\kilo\watt} & 30.36 & 15.68 & 1.10 & 52.55 & 0.66 & 1.67 & 0.84 \\
			AlTiBN-5 & Al$_{55}$Ti$_{35}$B$_{10}$ & 3\si{\kilo\watt} & 31.31 & 
			14.18 & 1.19 & 53.31 & 0.69 & 1.59 & 0.80 \\
			
			\addlinespace
			
			AlTiBN DC & Al$_{55}$Ti$_{35}$B$_{10}$ + Ti & 3\si{\kilo\watt} and
			1.0\si{\kilo\watt} & 27.87 & 17.67 & 3.68 & 50.01 & 0.61 & 3.51 & 1.76 \\

			\bottomrule

		\end{tabular}
	\end{adjustbox}
	\caption{Composition results extracted from GDOES analysis.}

	\label{tab:Composition}
\end{table}

\subsection{XRD diffraction}\label{subsec:XRD}

\begin{figure}
		\centering
		\includegraphics[width=0.45\textwidth]{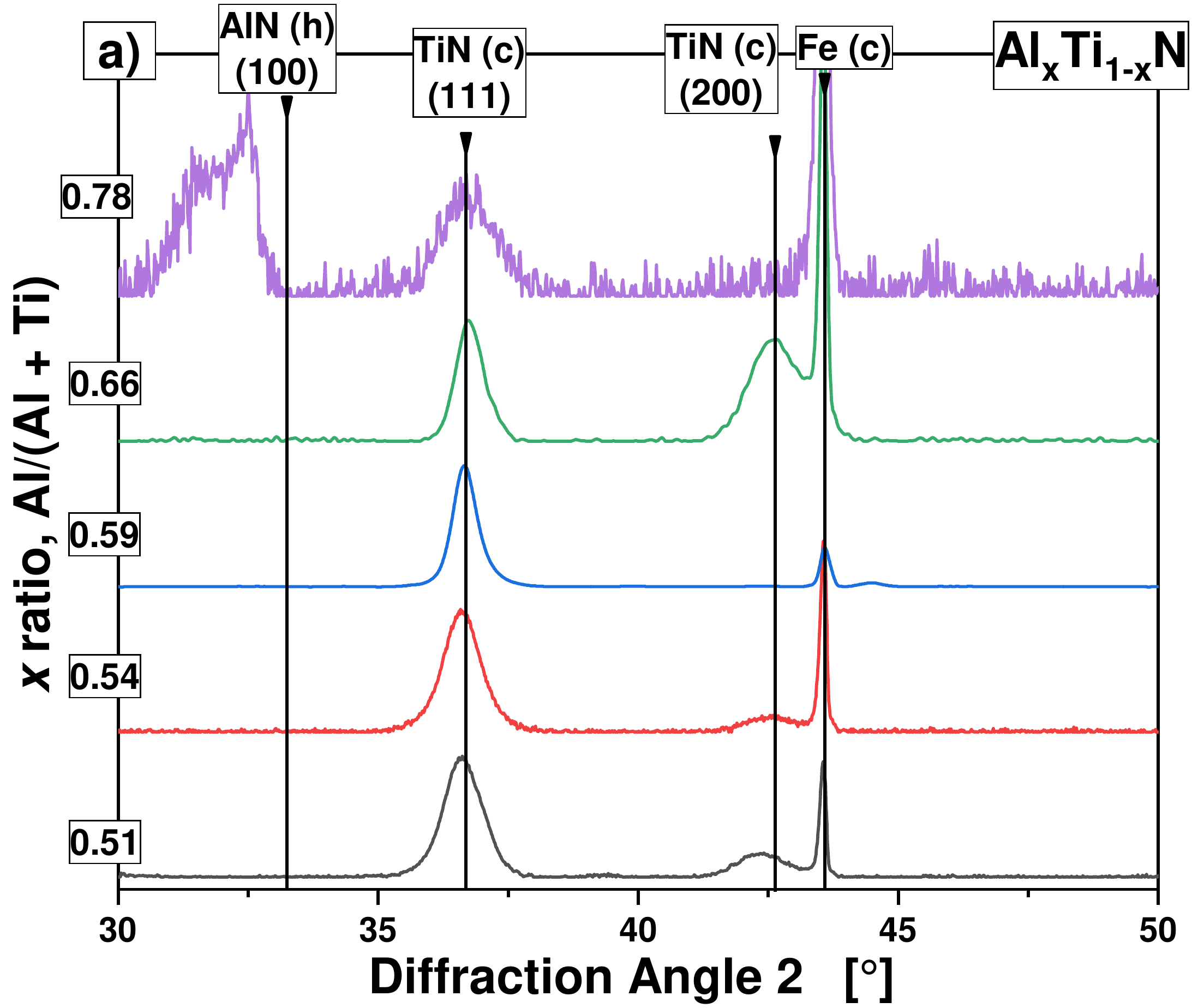}
		\includegraphics[width=0.45\textwidth]{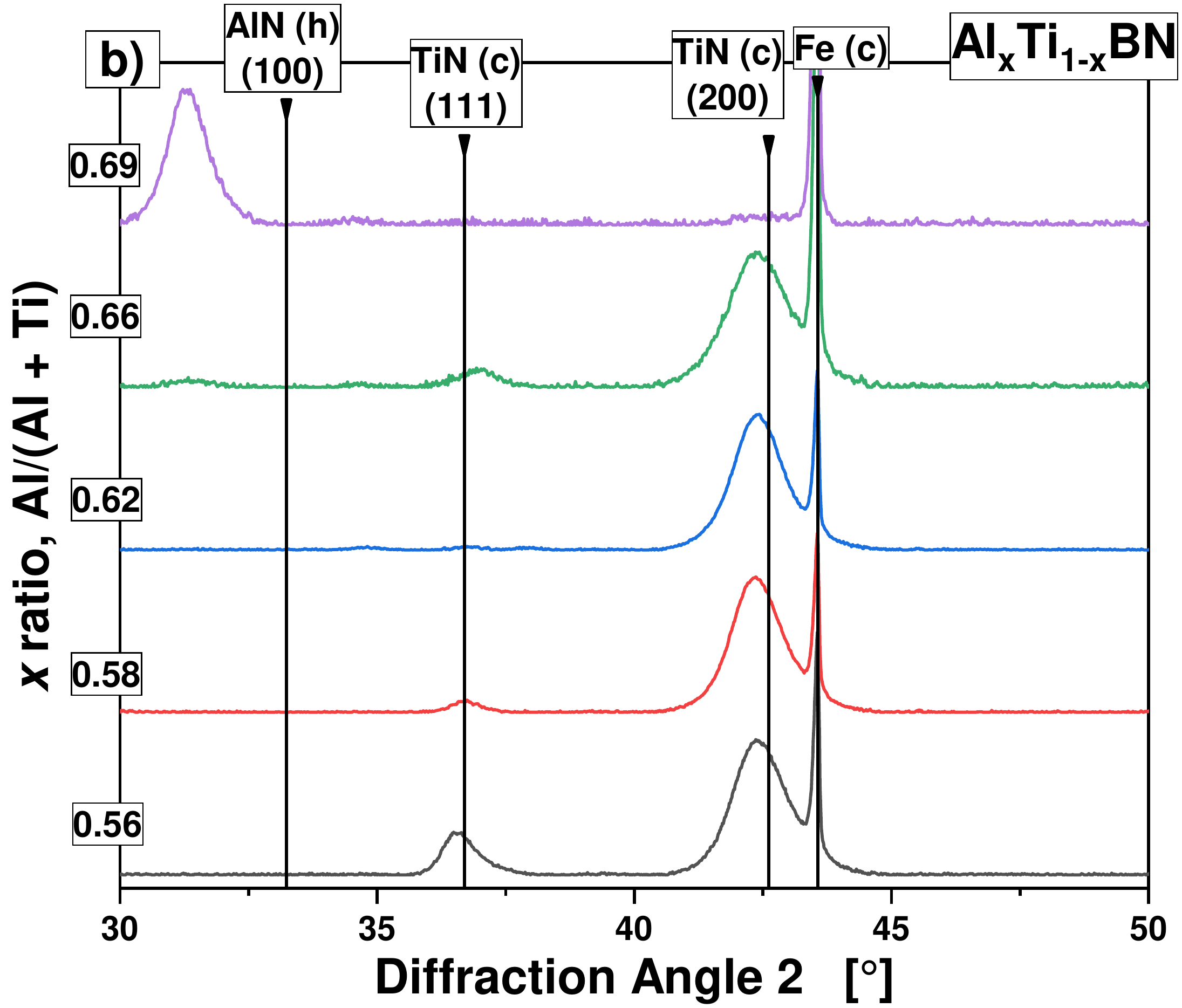}
		\includegraphics[width=0.5\textwidth]{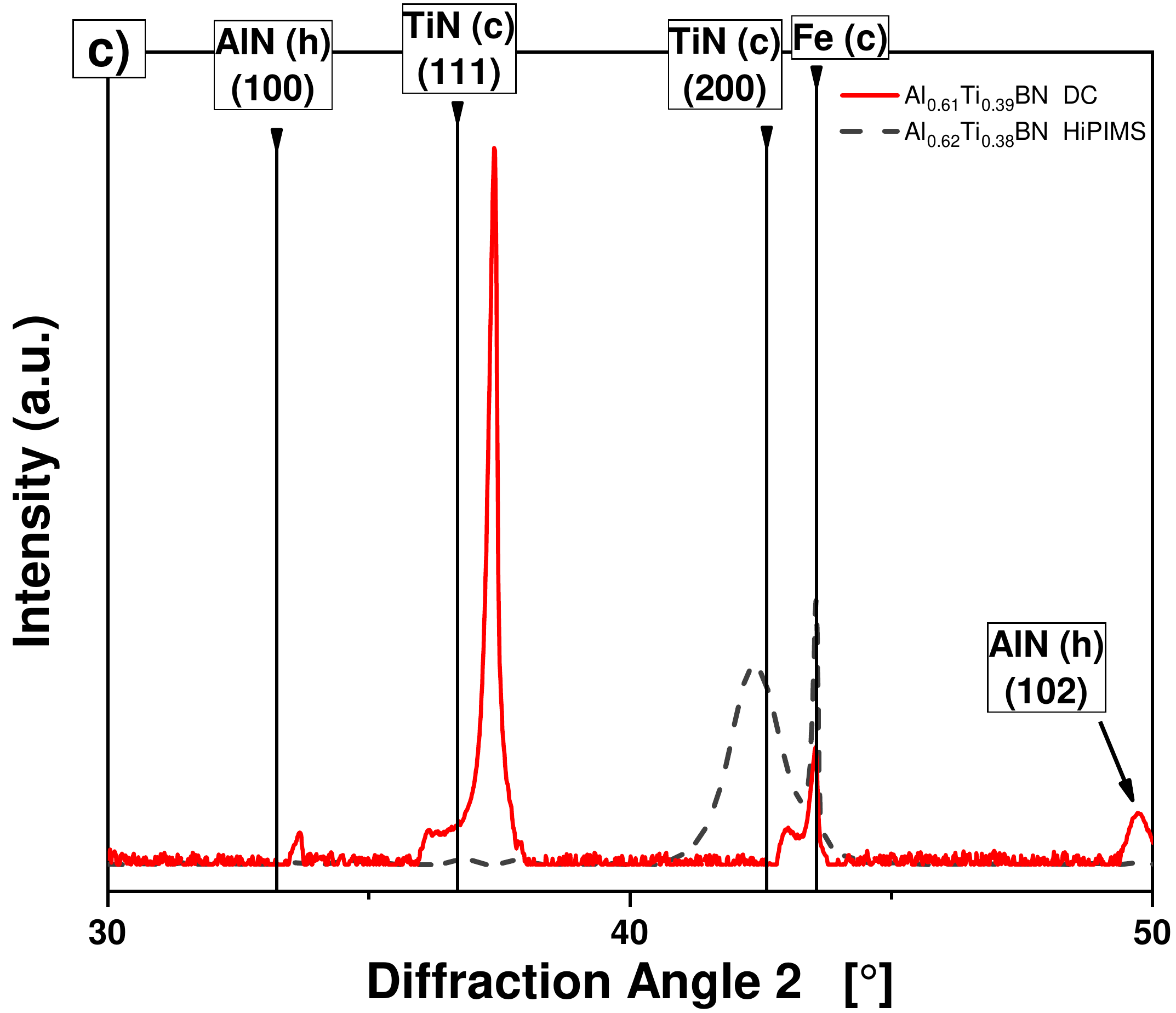}
		\caption{XRD patterns of as~-deposited (a) Al$_x$Ti$_{1-x}$N series and (b) Al$_x$Ti$_{1-x}$BN series, as a function of Al ratio $x$, and (c) the reference Al$_{0.61}$Ti$_{0.39}$BN DC in Bragg-Brentano configuration (square root of the intensity for clarity).}
	\label{fig:TiAlN_TiAlBN_BB}
\end{figure}

Fig. \ref{fig:TiAlN_TiAlBN_BB}(a) shows the XRD patterns for the AlTiN series as a function of $x$ ratio. The narrow peak at $2\theta \approx \SI{43.5}{\degree}$ corresponds to the steel substrate in all cases. Up to $x$ ratios of 0.66, the XRD patterns show that the coatings are composed by a single face centered cubic phase, which corresponds to the cubic $\delta$-Ti(Al)N rock-salt type B1 crystal structure \cite{tinpattern}. The diffraction peaks are relatively broad, presumably due to the high density of point and line defects induced by the highly energetic deposition conditions introduced by HiPIMS. Moreover, the intensity of the (111) peak was significantly larger than that of the (200) peak, indicating a preferred (111) growth orientation.  Finally, higher $x$ ratios, up to 0.78, resulted in a significant broadening of the (111) peak, which indicates a loss of crystallinity, and the appearance of a new diffraction peak, at $2\theta \approx \SI{33}{\degree}$, relatively close to the peak expected for hexagonal wurtzite-type B4 AlN phase \cite{alnhpattern}, but shifted to lower $2\theta$ angles. This is expected as the lattice parameter of wurtzite type Al$_x$Ti$_{1-x}$N has been reported to be slightly larger than for pure AlN, due to the larger size of the Ti atoms \cite{kimura2000}.

The XRD patterns for the AlTiBN series as a function of $x$ ratio are plotted in Fig. \ref{fig:TiAlN_TiAlBN_BB}(b). Similarly to the AlTiN series, the coatings are composed by a single crystalline phase, corresponding to the cubic $\delta$-Ti(Al)N rock-salt-type B1 crystal structure, for $x$ ratios up to 0.62. However, contrary to the AlTiN coatings grown under similar conditions, the AlTiBN coatings show a (200) preferred growth orientation. Again, for $x$ ratios higher than 0.66, a new diffraction peak, centered at $2\theta \approx \SI{32}{\degree}$ emerged, close to the peak expected for the hexagonal wurtzite-type phase, that becomes the dominant peak for an $x$ ratio of 0.69. 

Finally, the XRD pattern of the DC-sputtered Al$_{0.61}$Ti$_{0.39}$BN coating in Fig. \ref{fig:TiAlN_TiAlBN_BB}(c) reveals a single, very intense peak that corresponds to the (111) peak of $\delta$-Ti(Al)N, as opposed to the broad (002) peak found for the coating with a similar composition deposited by HiPIMS, Al$_{0.62}$Ti$_{0.38}$BN, which is over-imposed in the plot for comparison. Hence, the high ionization conditions introduced by HiPIMS clearly contribute to changes in the crystallinity and preferred growth orientation of the AlTi(B)N coatings.

\subsection{Coating morphology and microstructure}\label{subsec:MorphMicro}

\begin{figure}
	\centering
		\includegraphics[width=\textwidth]{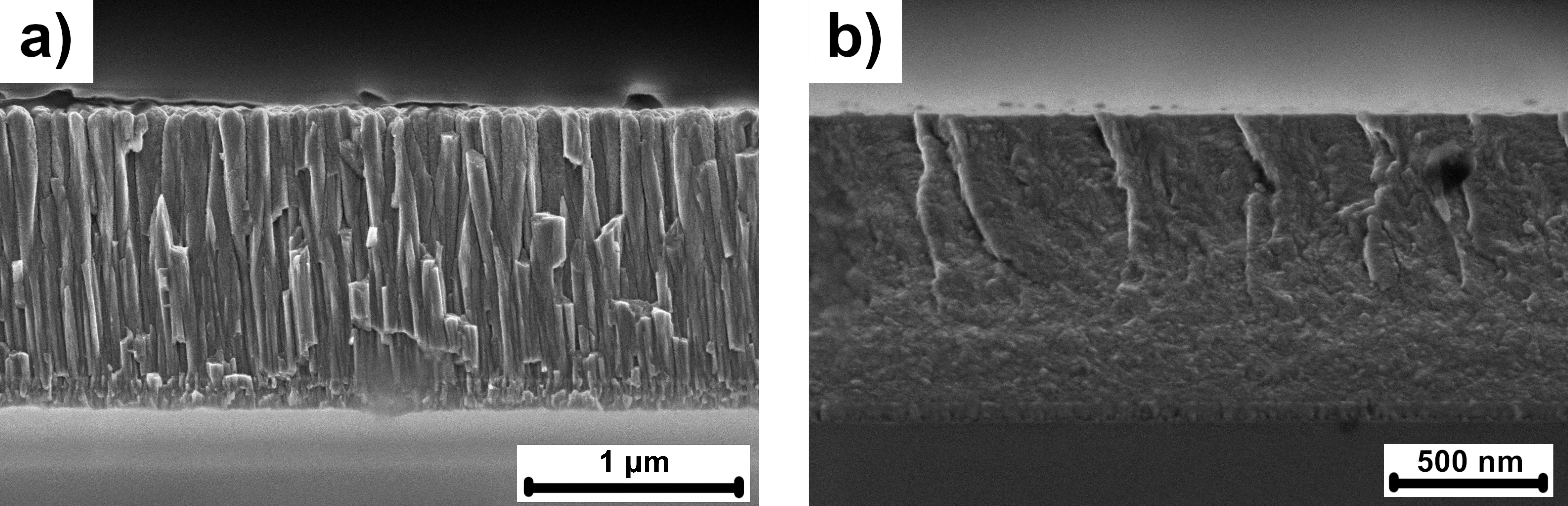} 
		\caption{SEM micrographs showing cross-sections from the (a) Al$_{0.59}$Ti$_{0.41}$N  and (b) Al$_{0.62}$Ti$_{0.38}$BN coatings. The first shows a columnar structure while B addition induces a fine-grained microstructure.}
	\label{fig:SEM_Comparison}
\end{figure}

The XRD studies revealed that B addition introduced significant changes in the preferred growth orientation of the coatings. In order to assess its effect on the coating morphology and microstructure, one coating of each series, Al$_{0.59}$Ti$_{0.41}$N (AlTiN-3) and Al$_{0.62}$Ti$_{0.38}$BN (AlTiBN-3), both with similar $x$ ratios, were selected for further microstructural analysis. Fig. \ref{fig:SEM_Comparison} shows SEM images of the fractured cross section of each coating. The Al$_{0.59}$Ti$_{0.41}$N coating in Fig. \ref{fig:SEM_Comparison}(a), with a preferred (111) growth orientation, shows a clear columnar morphology, characteristic of sputtered coatings. In contrast, the Al$_{0.62}$Ti$_{0.38}$BN coating in Fig. \ref{fig:SEM_Comparison}(b) presents a clearly different morphology, with no signs of columnar grains, which might be indicative of a fine-grained microstructure.

\begin{figure}
	\centering
		\includegraphics[width=\textwidth]{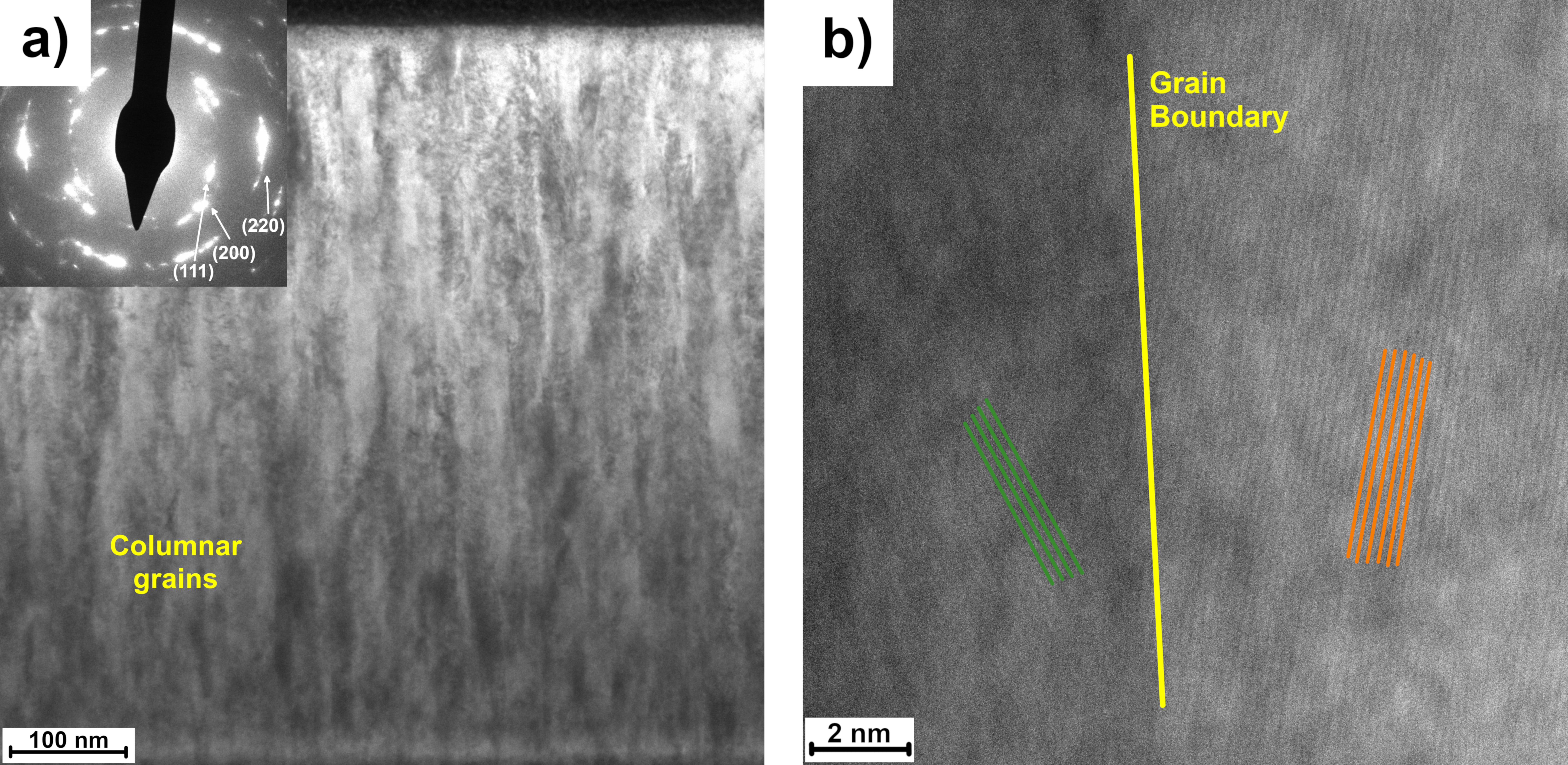} 
		\caption{a) Cross-sectional BF-TEM image of the Al$_{0.59}$Ti$_{0.41}$N coating, showing a columnar grain structure oriented in the growth direction (upwards), typical of sputtered coatings. The insert corresponds to the DP from the same area. b) HR STEM image in the vicinity of a grain boundary. Lattice fringes within each grain are marked by parallel lines, and are continuous up to the grain boundary, with no evidence for the presence of any amorphous phase. Green-marked lattice fringes were assigned to (111) crystallographic planes, while orange-marked lattice fringes were assigned to (200) crystallographic planes, based on their interplanar spacing (see text).} 
	\label{fig:TiAlN_Combined}
\end{figure}

Further microstructural details of the same coatings were obtained through cross~-sectional TEM analysis, by extracting two electron transparent lamellae using the lift-out technique in a FIB-FEGSEM. Fig. \ref{fig:TiAlN_Combined}(a) shows a bright-field (BF) image of the Al$_{0.59}$Ti$_{0.41}$N (AlTiN-3) coating that reveals \SI{\sim20}{\nano\meter} wide columnar grains oriented in the growth direction, confirming what was seen in the SEM. The insert in Fig. \ref{fig:TiAlN_Combined}(a) corresponds to the diffraction pattern (DP) from the same area confirming a (111) preferred growth orientation, in agreement with the XRD results. The coating is fully crystalline, as shown in Fig. \ref{fig:TiAlN_Combined}(b), which corresponds to a high-resolution scanning transmission electron microscopy (HR STEM) image, in high-angle annular dark~-field (HAADF) mode, taken around a particular columnar grain boundary. The image reveals that the lattice planes extend continuously in each of the columnar grains and are interrupted by the grain boundary, with no evidence for the presence of any amorphous phase. Measured spacing between green-marked lattice fringes in Fig. \ref{fig:TiAlN_Combined}(b) was \SI{2.46}{\angstrom}, corresponding to the (111) crystallographic planes; while the measured spacing between the orange-marked lattice fringes was \SI{2.10}{\angstrom}, corresponding to the (200) crystallographic planes.

\begin{figure}
	\centering
		\includegraphics[width=\textwidth]{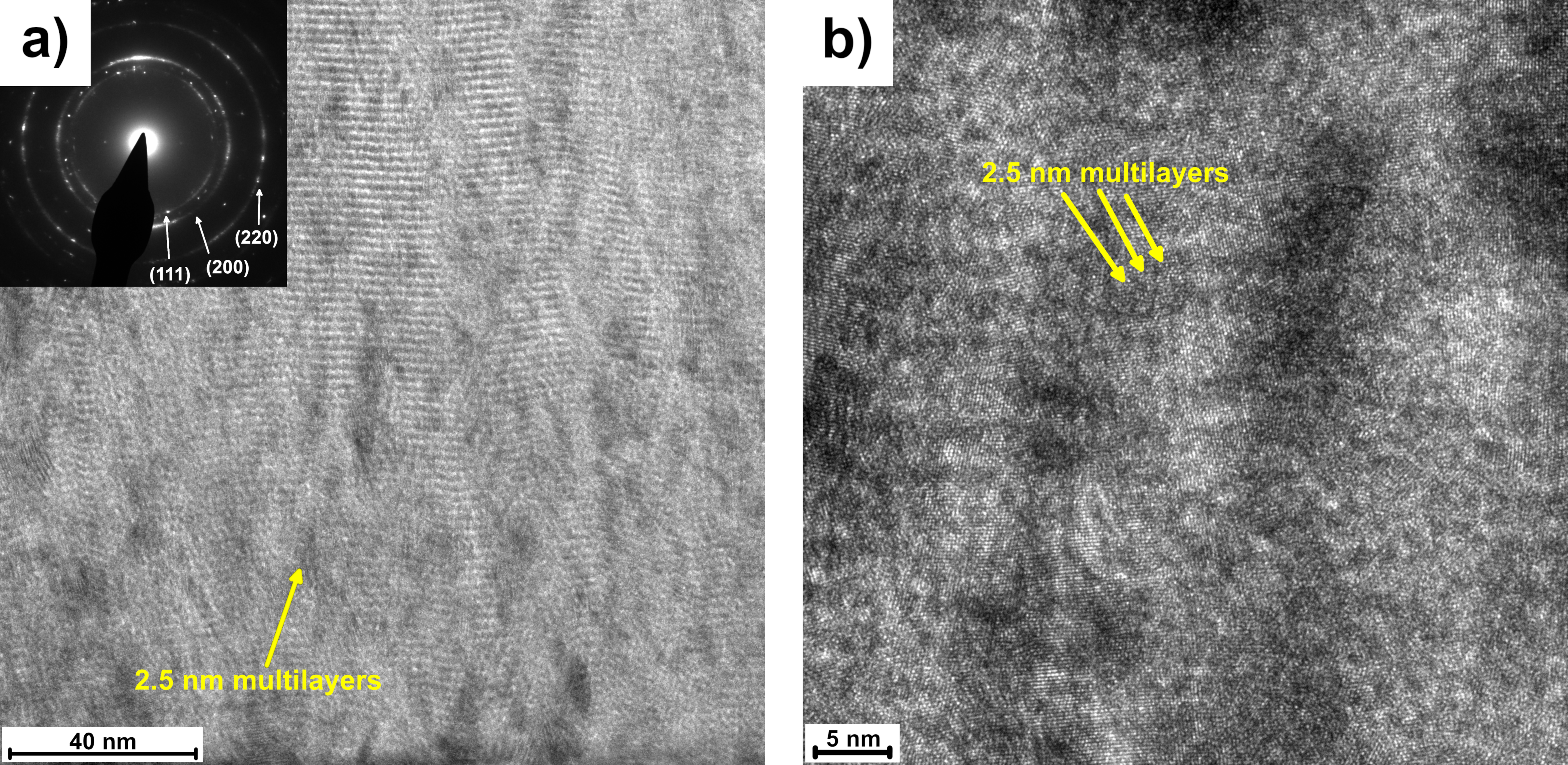} 
		\caption{a) Cross-sectional BF-TEM image of Al$_{0.62}$Ti$_{0.38}$BN coating, showing a fine nanograined multilayered structure; b) HREM image showing nanocrystalline and amorphous domains. The growth direction is vertical in both cases.}
	\label{fig:TiAlBN_Combined}
\end{figure}

\begin{figure}
	\centering
		\includegraphics[width=\textwidth]{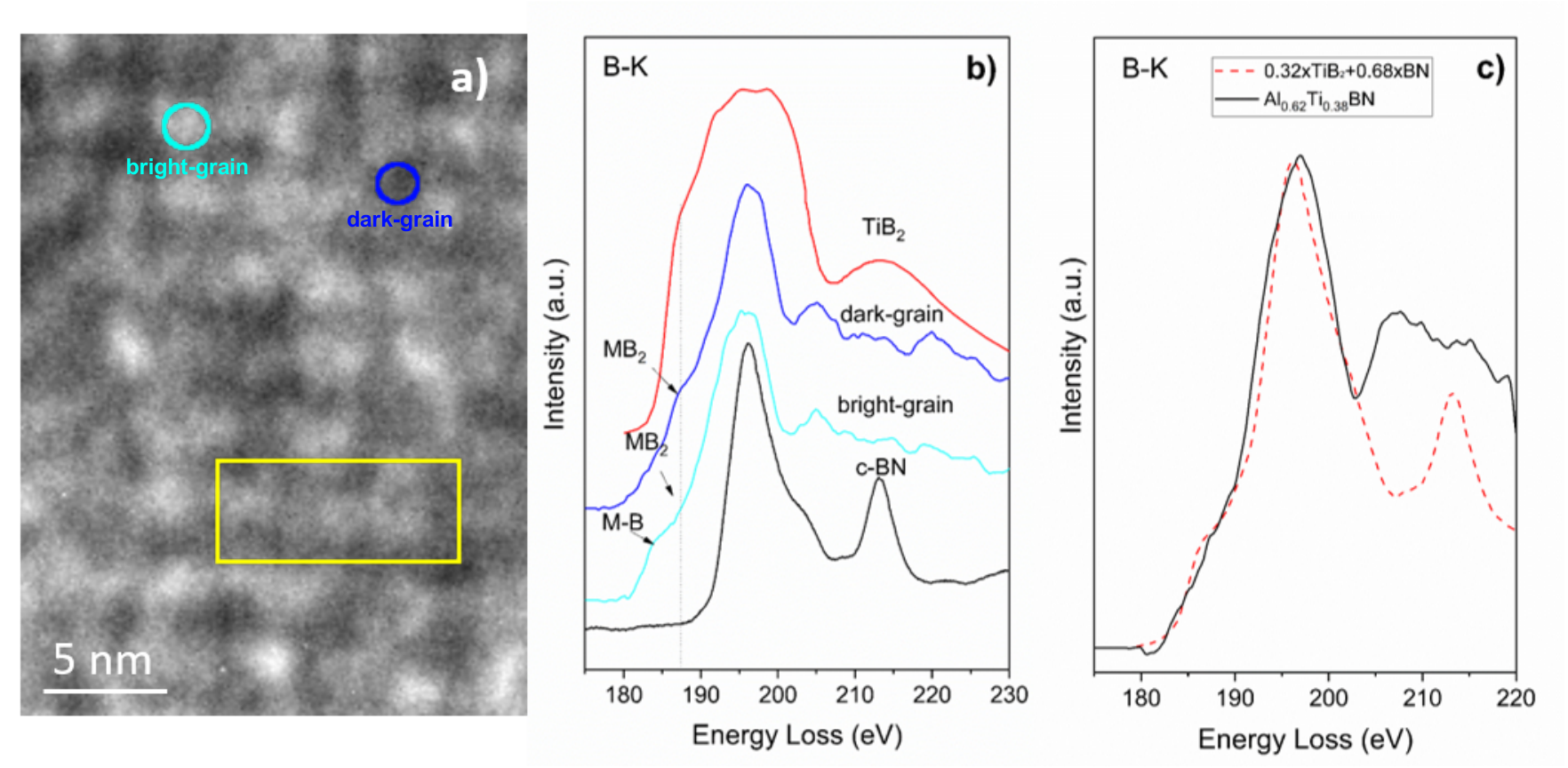} 
		\caption{a) High magnification HAADF-STEM or Z-contrast image of sample Al$_{0.62}$Ti$_{0.38}$BN, b) B$-K$ edge EELS spectra measured in the dark and bright grains of the coating, compared with reference TiB$_2$, and c-BN. c) Spectra obtained from the lineal combination of TiB$_2$ and BN, that fit with the integrated B$-K$ edge from the yellow area of the figure (see the text).}
	\label{fig:EELS}
\end{figure}

On the other hand, the BF-TEM image of the Al$_{0.62}$Ti$_{0.38}$BN coating (AlTiBN-3) in Fig. \ref{fig:TiAlBN_Combined}(a) revealed a completely different microstructure. In agreement with the SEM observations, the microstructure was not composed of columnar grains, but the addition of B triggered the formation of a more nanocrystalline microstructure. This was clearly shown in the DP included as insert in Fig. \ref{fig:TiAlBN_Combined}(a), which, contrary to the Al$_{0.59}$Ti$_{0.41}$N coating, was composed of continuous diffraction rings instead of diffraction spots, and in the HAADF-STEM image of Fig. \ref{fig:EELS}(a). Moreover, the (200) diffraction ring was more intense in the vertical growth direction, confirming the (200) preferred growth orientation determined by XRD. The TEM image of Fig. \ref{fig:TiAlBN_Combined}(a) also indicated the formation of a multilayered structure, with an individual layer thickness of the order of \SI{2.5}{\nano\meter}. The layered structure arises as a consequence of the rotation of the substrate during co-deposition from the two targets, Al$_{55}$Ti$_{35}$B$_{10}$ and Ti. This was confirmed by depositing an extra coating (not shown here), using the same deposition conditions, but with a reduced substrate rotation speed, which resulted in a larger layer thickness. Therefore, the layered structure reflects a chemical modulation within the coating between nitride layers that are richer in Al and B and layers that are richer in Ti. It was not possible to confirm such chemical modulation by energy dispersive spectroscopy (EDS), due to limitations in the resolution of this technique, but further insights from EELS spectroscopy are given below. Interestingly, the Al$_{0.59}$Ti$_{0.41}$N coating in Fig. \ref{fig:TiAlN_Combined}(a) did not show any evidence of a multilayered structure, despite being also co-deposited from two targets, Al$_{50}$Ti$_{50}$ and Al, at the same substrate rotation speed. Therefore, the presence of B seems to be crucial for the formation of the multilayered structure, as will be further discussed below. As a matter of fact, the HREM image of Fig. \ref{fig:TiAlBN_Combined}(b) revealed that, in addition to the layered structure, the microstructure of Al$_{0.62}$Ti$_{0.38}$BN was composed of a mixture of nanocrystalline and amorphous domains. These domains have different composition, as shown in the high magnification HAADF-STEM or Z-contrast image of Fig. \ref{fig:EELS}(a), where dark and bright nanograins, with a diameter around \SI{2}{\nano\meter}, can be observed. 

EELS spectroscopy is a suitable technique to study the presence of amorphous phases and to determine the boron content and its chemical bonding \cite{arzac2014} with a lateral resolution of the order of the electron probe size, which was around \SI{\sim1}{\nano\meter} in this work. For this, the B$-K$, N$-K$, Ti $L_{2,3}$ and Al$-K$ edges were measured in different locations of the Al$_{0.62}$Ti$_{0.38}$BN coating. Fig. \ref{fig:EELS}(b) shows the B$-K$ edge recorded in dark and bright nanograins indicated in Fig. \ref{fig:EELS}(a). In order to elucidate the chemical bonding of the B atoms, the recorded EELS spectra were compared with reference spectra for c-BN and TiB$_2$. Since TiB$_2$ and AlB$_2$ are isostructural compounds, they are expected to display similar spectra \cite{lie2000}. The measured spectra are quite comparable to the c-BN reference, but the widening of the main resonance (at \SI{195}{\electronvolt}) and the presence of a shoulder at lower energy loss (\SI{188}{\electronvolt}), labeled MB$_2$ in the figure, gives account for the presence of Ti(Al)B$_2$. The only difference between the spectra recorded in the dark and bright nanograins is that bright grains display a small peak a lower energy loss (\SI{182}{\electronvolt}), that can be assigned to the presence of Ti(Al)$-$B bonds, labeled as MB in the figure, which might originate from B atoms either incorporated into the (Ti,Al)N grains or located at the grain boundaries \cite{baker2002, baker2007}. Apart from this, the N$-K$, Ti $L_{2,3}$ and Al$-K$ edges, not shown here, were identical in the bright and dark nanograins and compatible with the formation of (Ti,Al)N, c-BN and Ti(Al)B$_2$, as expected from the (Ti,Al)$-$B$-$N equilibrium phase diagram \cite{baker2002}.

Overall, the EELS spectra recorded in the bright and dark nanograins showed no evidence for the strong chemical modulation suggested by the multilayered structure seen in the HAADF-STEM image of Fig. \ref{fig:EELS}(b). This might be due to the fact that, despite the small diameter of the electron probe (\SI{\sim1}{\nano\meter}), the measurements account for the average electron energy loss along the thickness of the TEM foil, which is of the order of \SI{100}{\nano\meter}, and therefore several layers and nanograins might overlap along its thickness, due to the layer waviness. Because of this, the quantitative analysis of the EELS spectra was performed based on the average of all measurements, and considering that EELS is not particularly well suited for the quantification of heavier elements like Al or Ti, the chemical composition was estimated by fixing the $(Ti+Al)/N$ ratio from the GDOES measurements (0.9 from Table \ref{tab:Composition}) and the $B/N$ ratio obtained by EELS, 0.33. Expressing the atomic ratios per unit of metallic atom (Al or Ti), \textit{i.e.} (Ti$_{1-x}$Al$_x$)B$_y$N$_z$, results in an stoichiometry for this particular coating of (Ti$_{0.42}$Al$_{0.58}$)B$_{0.36}$N$_{1.1}$ and a B content of approximately \SI{14}{\atpercent}, which agrees well with the expected B content from the composition of the Al$_{55}$Ti$_{35}$B$_{10}$ target. 
Therefore, the results suggest that the coating is composed on nanocrystalline domains of a Ti(Al)N phase, of size around \SI{2}{\nano\meter}, surrounded by amorphous domains of BN and Ti(Al)B$_2$. Following \cite{baker2007}, it is possible to estimate the average percentage of each phase from:

\begin{equation}
C_{Ti(Al)B_2}=\frac{1+y-z}{1+y+2z}
\end{equation}

\begin{equation}
C_{Ti(Al)N}=\frac{2-y+z}{1+y+2z}
\end{equation}

\begin{equation}
C_{BN}=\frac{-2+y+2z}{1+y+2z}
\end{equation}

Obtaining 77\% Ti(Al)N, 7.3\% Ti(Al)B$_2$ and 15.7\% BN. With these TiB$_2$ and BN percentages, we obtain a B$-K$ spectra that fit quite well with the EELS spectra integrated over the marked area of Fig. \ref{fig:EELS}(c). Since the mutilayered structure arises from the substrate rotation exposing the coating alternatively to the Ti and Al$_{55}$Ti$_{35}$B$_{10}$ targets, the bright layers should be richer in the crystalline Ti(Al)N phase, while the dark layers should contain more amorphous BN and Ti(Al)B$_2$ phases. Unfortunately, the EELS measurements did not show clear evidence of the latter due to the small thickness of the layers.

\subsection{Mechanical Properties}\label{subsec:Mech}

\begin{figure}
	\centering
		\includegraphics[width=0.8\textwidth]{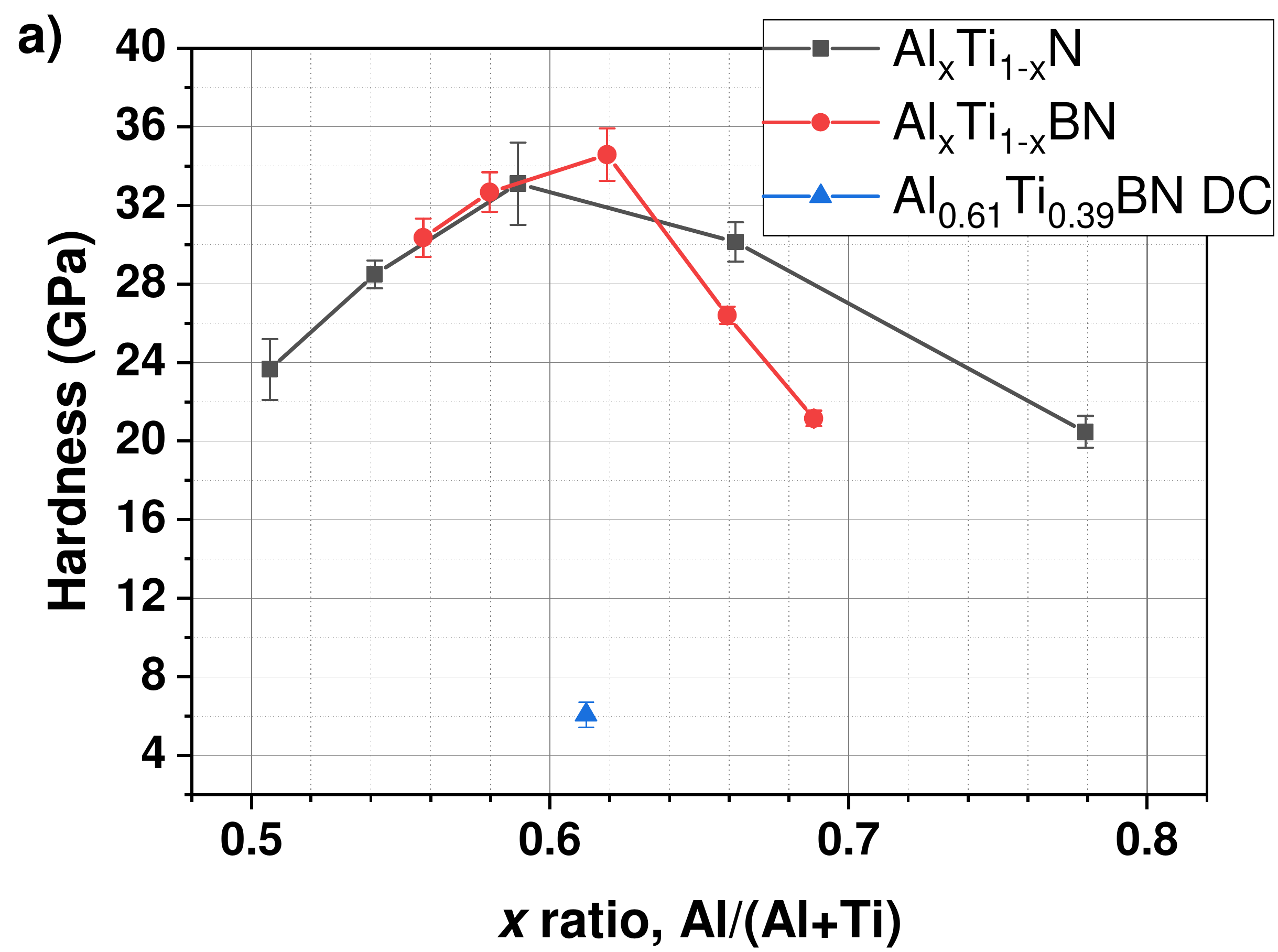} 
		\includegraphics[width=0.8\textwidth]{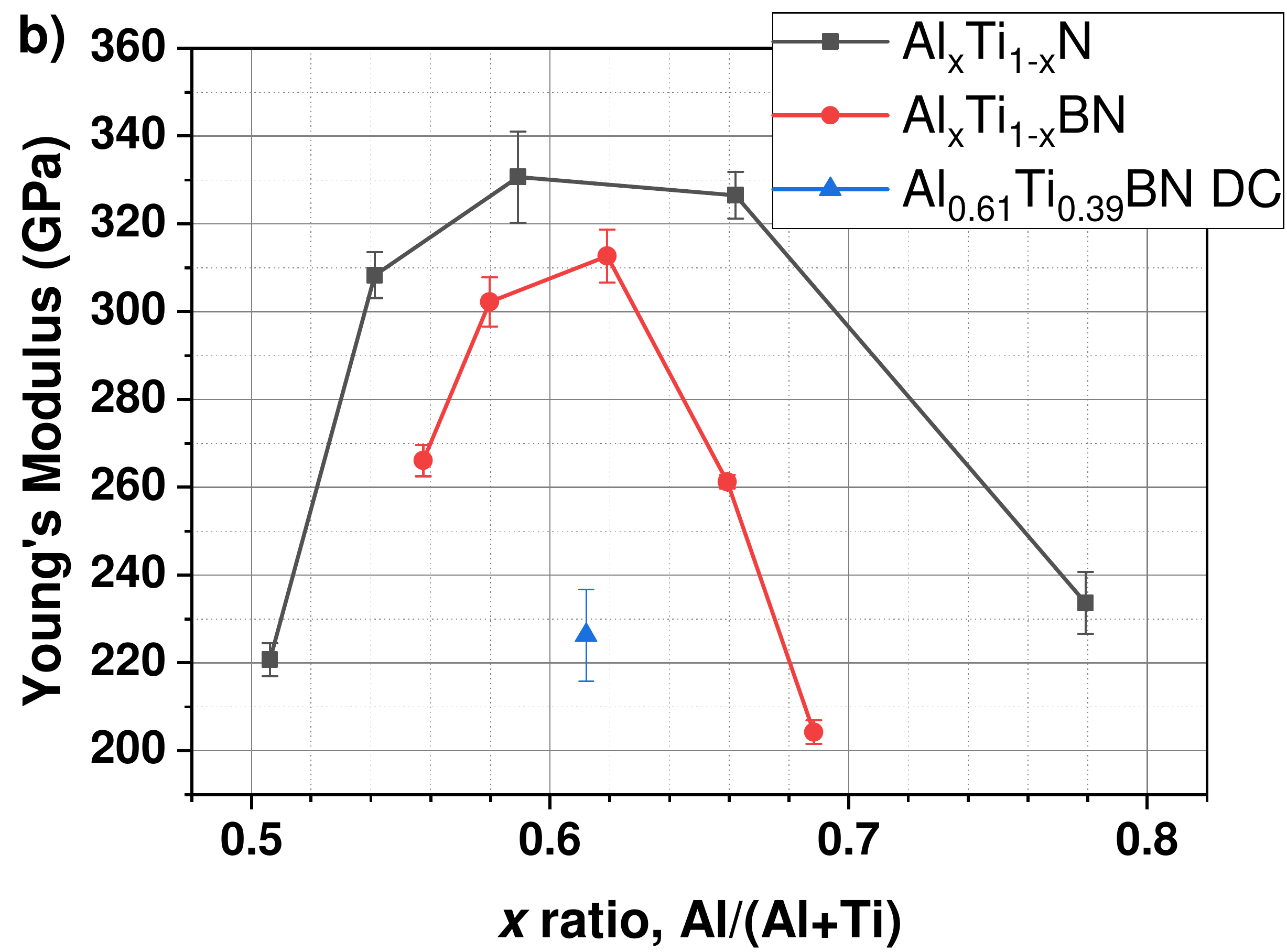}
		\caption{Evolution of (a) hardness and (b) elastic modulus \textit{versus} $x$ ratio for the Al$_x$Ti$_{1-x}$N and Al$_x$Ti$_{1-x}$BN series. Both series show the maximum hardness for $x$ ratios around 0.6.}
	\label{fig:MechProps}
\end{figure}

Fig. \ref{fig:MechProps}(a) plots the hardness (H) and elastic modulus (E) of the as-deposited Al$_x$Ti$_{1-x}$N and Al$_x$Ti$_{1-x}$BN coatings as a function of $x$ ratio. Both series followed a similar trend, reaching a maximum hardness in both cases for $x$ ratios around 0.6. The Al$_x$Ti$_{1-x}$N coatings reached the maximum hardness, \SI{33}{\giga\pascal}, for $x=0.58$, while the Al$_x$Ti$_{1-x}$BN coatings reached the maximum, \SI{35}{\giga\pascal}, for $x=0.62$. As shown in Fig. \ref{fig:MechProps}(a), the hardness was considerably higher than the one obtained in the Al$_{0.61}$Ti$_{0.39}$BN coating deposited in DC (\SI{5}{\giga\pascal}), demonstrating the benefit of the high ion bombardment conditions achieved by the use of HiPIMS. 

\begin{figure}
	\centering
		\includegraphics[width=0.45\textwidth]{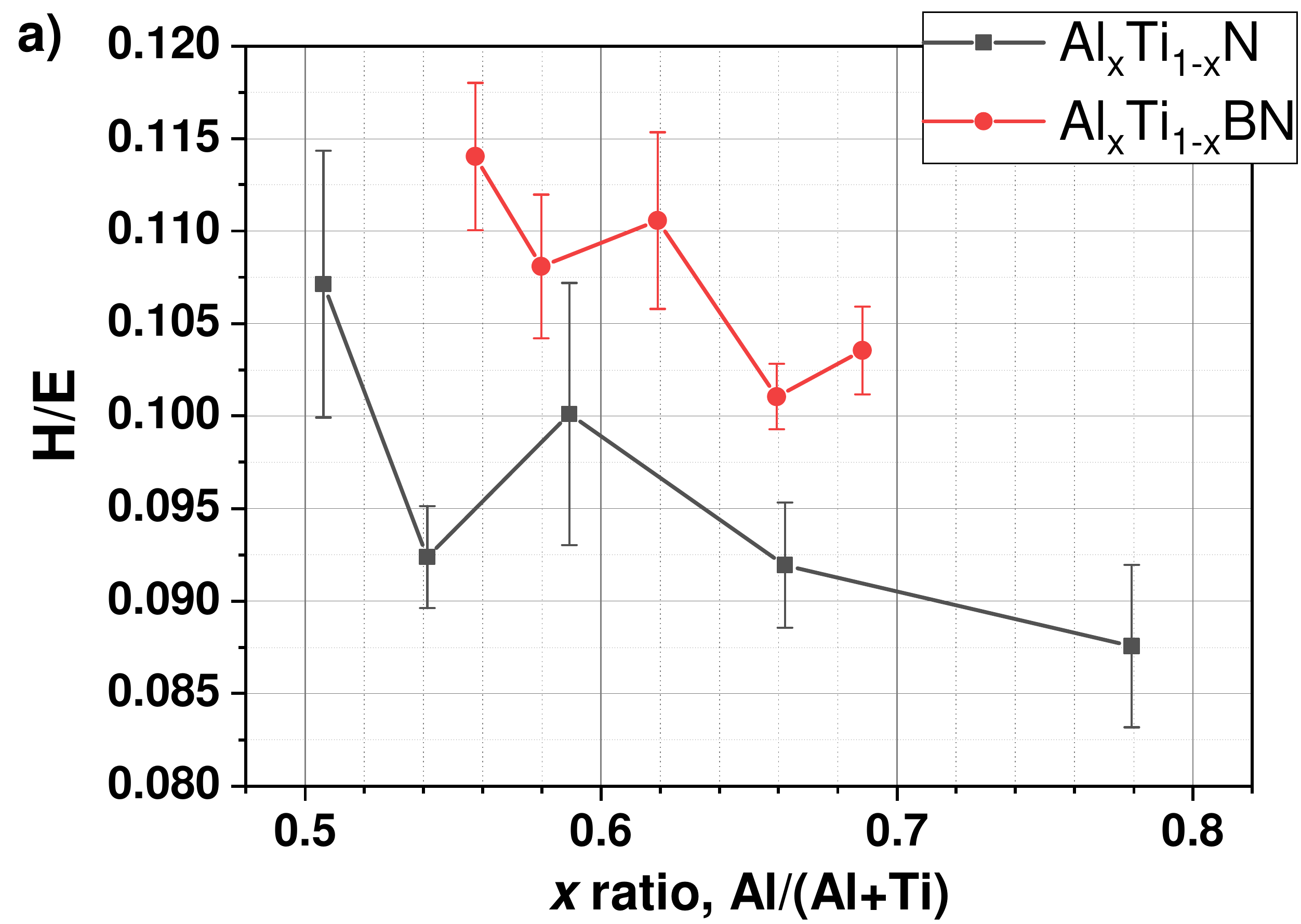} 
		\includegraphics[width=0.45\textwidth]{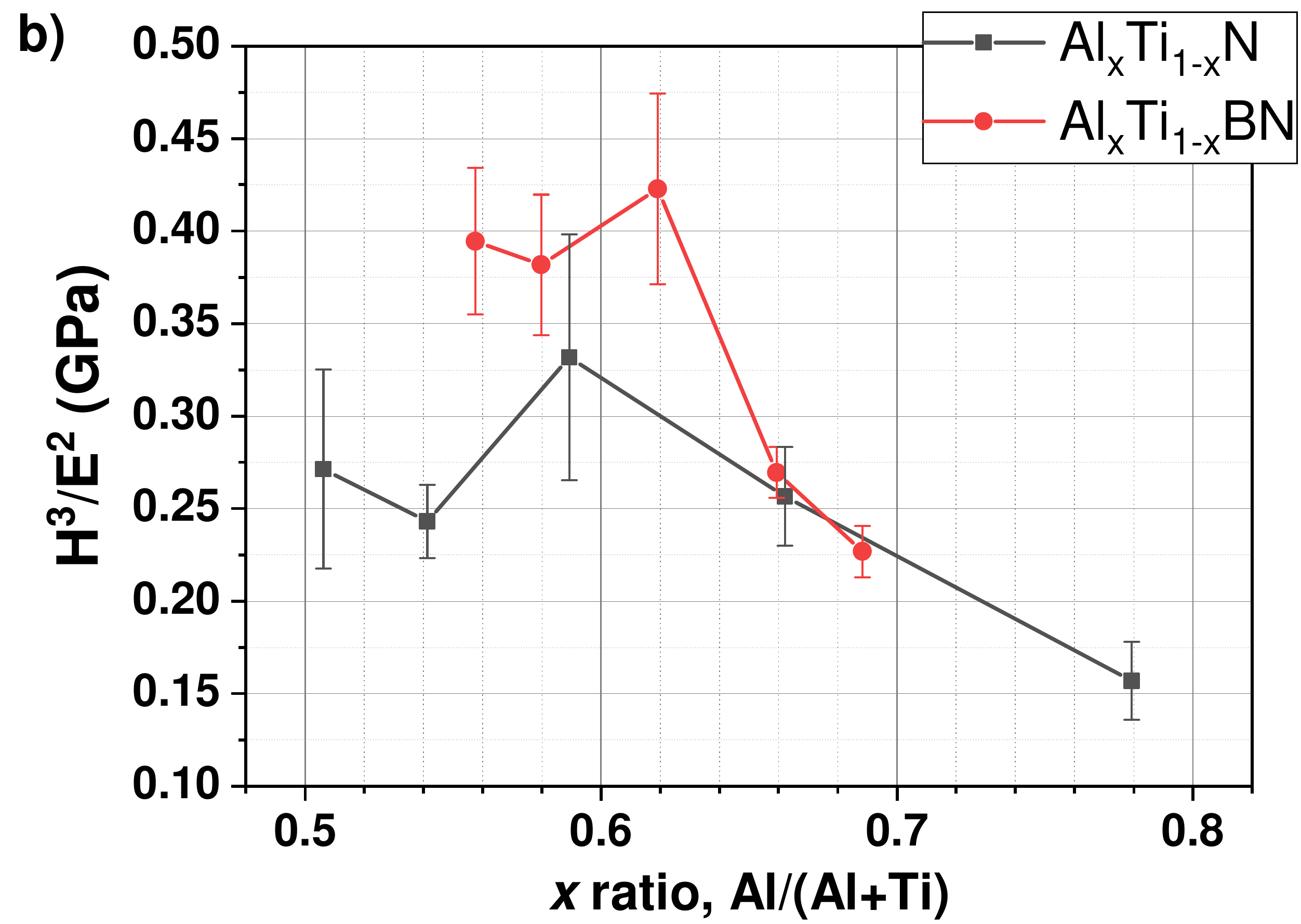}
		\caption{a) Plot of the $H/E$ ratio \textit{versus} $x$ ratio. B-addition makes Al$_x$Ti$_{1-x}$BN more flexible than Al$_x$Ti$_{1-x}$N. b) Plot of the $H^3/E^2$ ratio \textit{versus} $x$ ratio. This value gives an estimate of fracture resistance, rendering Al$_x$Ti$_{1-x}$BN more fracture resistant than Al$_x$Ti$_{1-x}$N coatings in all cases.}
	\label{fig:HE_H3E2}
\end{figure}

Fig. \ref{fig:MechProps}(b) plots the elastic modulus as a function of $x$ ratio for the Al$_x$Ti$_{1-x}$N and the Al$_x$Ti$_{1-x}$BN series. The Al$_x$Ti$_{1-x}$BN coatings show lower elastic moduli than the Al$_x$Ti$_{1-x}$N coatings for all $x$ ratios. This is very relevant, as the $H/E$ and $H^3/E^2$ ratios are typically used to predict the tribological performance of coatings \cite{wood2015}. While, the $H/E$ ratio has been considered to correlate with the elastic strain to failure, the $H^3/E^2$ ratio has been proposed as a measure of the resistance to fracture. The changes in the $H/E$ and the $H^3/E^2$ ratios with $x$ ratio are plotted in Fig. \ref{fig:HE_H3E2}. The Al$_x$Ti$_{1-x}$BN coatings showed higher $H/E$ ratios than the Al$_x$Ti$_{1-x}$N coatings for all $x$ ratios. The optimum $x$ ratio in both coatings series was found to be around 0.6, as the maximum $H^3/E^2$ ratios in each series were 0.42 and 0.33 for Al$_{0.59}$Ti$_{0.41}$N and Al$_{0.62}$Ti$_{0.38}$BN, respectively.

\subsection{Thermal stability and hot hardness}\label{subsec:Annealing}

The best coatings in each series (Al$_{0.59}$Ti$_{0.41}$N and Al$_{0.62}$Ti$_{0.38}$BN) were subjected to annealing treatments to assess their thermal stability. The Al$_{0.62}$Ti$_{0.38}$BN coating went through annealing treatments up to \SI{1100}{\celsius}, while the Al$_{0.59}$Ti$_{0.41}$N coating could only be tested up to \SI{900}{\celsius}, because higher temperatures caused coating delamination.

\begin{figure}
	\centering
		\includegraphics[width=0.7\textwidth]{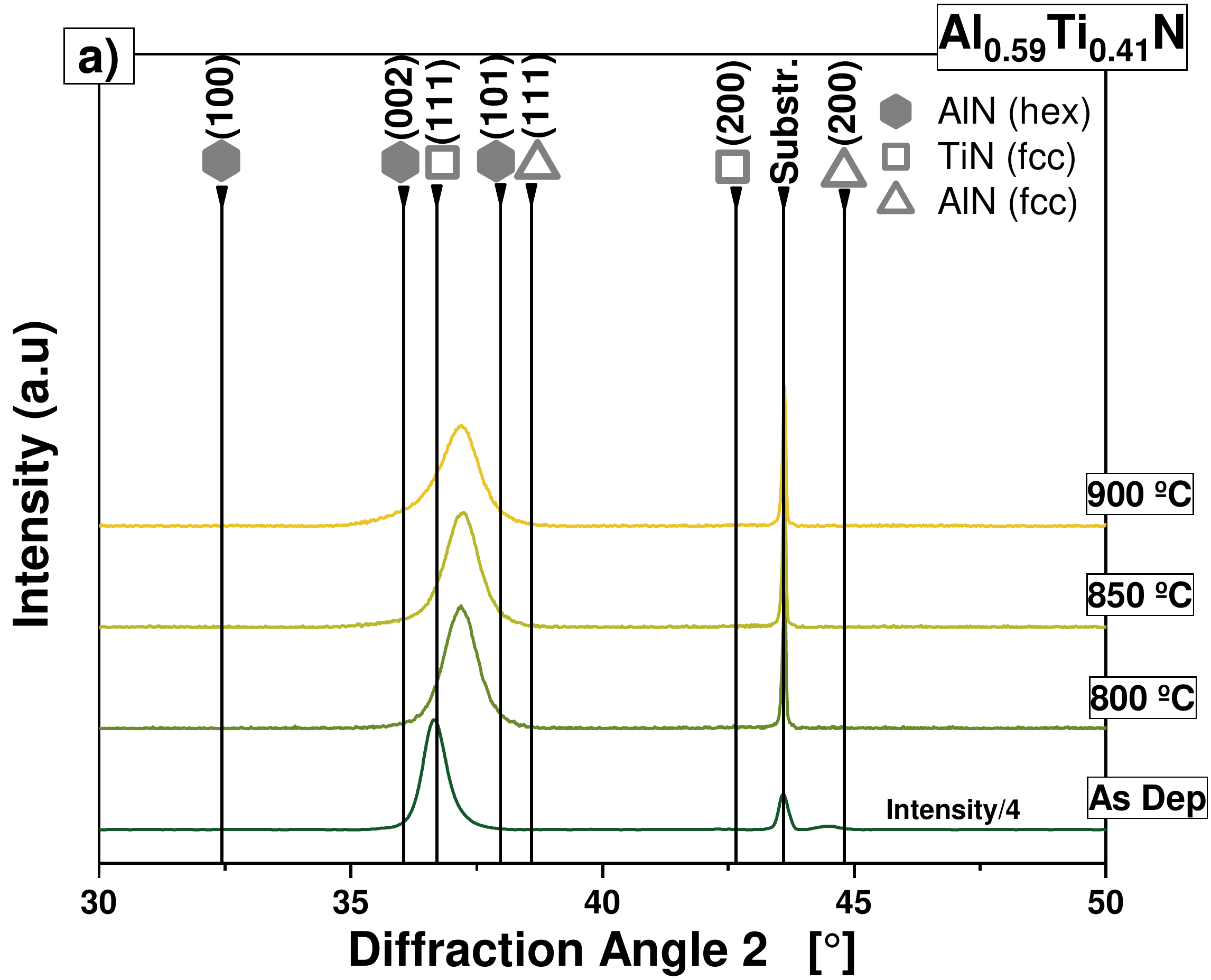} 
		\includegraphics[width=0.7\textwidth]{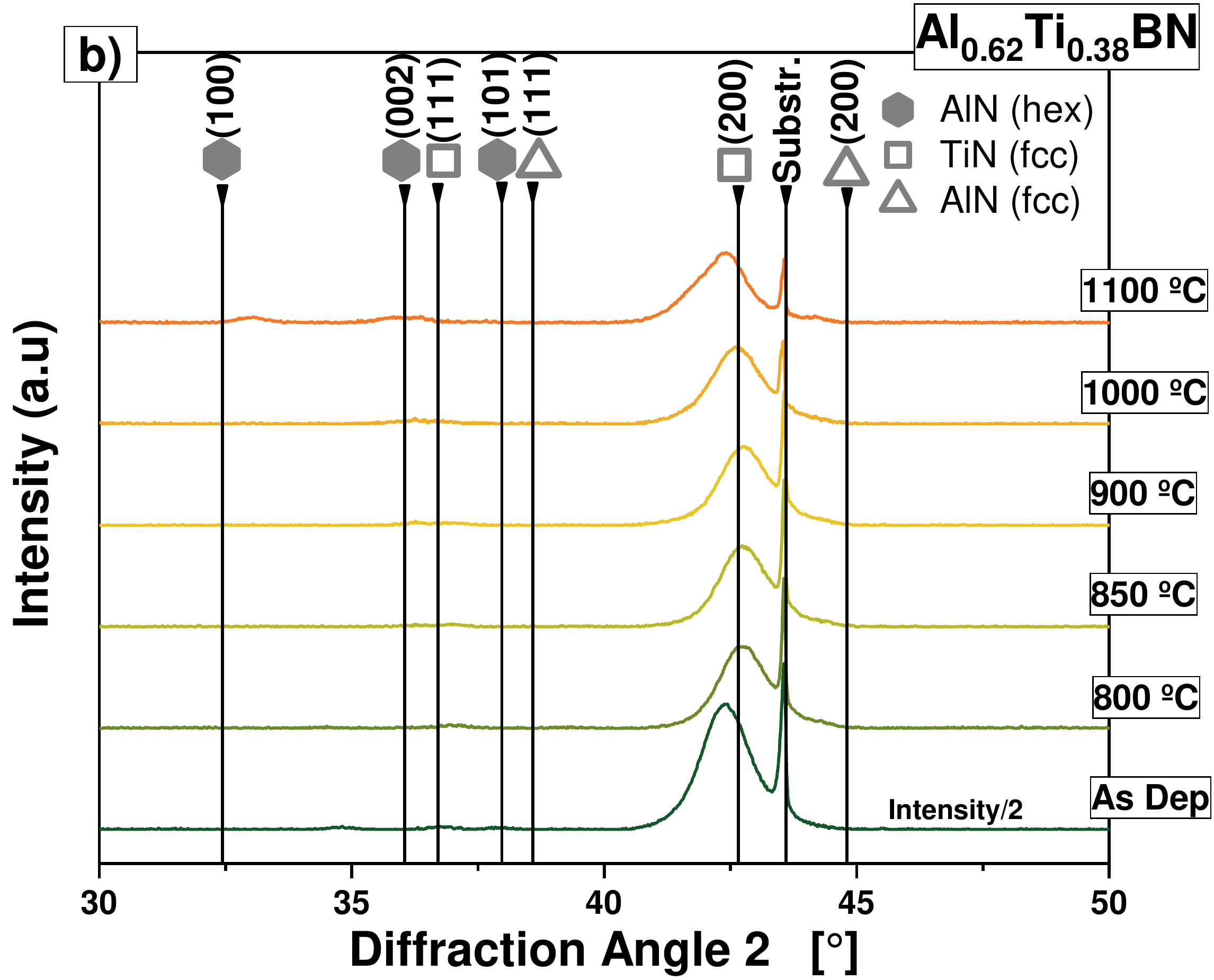}
		\caption{XRD patterns corresponding to the a) Al$_{0.59}$Ti$_{0.41}$N and b) Al$_{0.62}$Ti$_{0.38}$BN coatings after annealing at different temperatures.}
	\label{fig:XRD_Annealing}
\end{figure}

The XRD patterns of the Al$_{0.59}$Ti$_{0.41}$N and Al$_{0.62}$Ti$_{0.38}$BN coatings, as a function of annealing temperature, are presented in Figs. \ref{fig:XRD_Annealing}(a) and \ref{fig:XRD_Annealing}(b), respectively. In the case of the Al$_{0.59}$Ti$_{0.41}$N coating, the main changes upon annealing were found in the position and shape of the main diffraction peak corresponding to the (111) preferred growth orientation of the cubic nitride phase. In the as-deposited condition, the position of the (111) peak matched the theoretical position for stress-free pure $\delta$-TiN. This is surprising because this coating had a fully crystalline columnar microstructure (see Fig. \ref{fig:TiAlN_Combined}) and, for an $x$ ratio of 0.59, the (111) peak position should lay at intermediate positions between the theoretical $\delta$-TiN and $\delta$-AlN cubic phases. Its actual position at lower diffraction angles might have been the result of the compressive residual stresses induced by the high degree of ion bombardment expected from the HiPIMS conditions, as corroborated by the residual stress measurements presented in Sec. \ref{subsec:Rate}. As a matter of fact, upon annealing at \SI{800}{\celsius}, the (111) peak shifts to higher diffraction angles and its position agrees with the theoretical position for a stress-free solid solution Al$_{0.59}$Ti$_{0.41}$N cubic phase, presumably due to the recovery of built-in structural point and line defects and the relief of the compressive residual stresses. The peak shift to higher $2\theta$ angles also suggests phase separation and the formation of B1 cubic AlN domains (its lattice parameter is smaller than for TiN \cite{mayrhofer2006}). At even higher annealing temperatures, \SI{850}{\celsius}, the peak shifted back and showed a tendency to broaden asymmetrically towards lower diffraction angles. This has been shown to indicate the formation of hexagonal AlN \cite{bartosik2017}. Similar trends were observed in the case of the Al$_{0.62}$Ti$_{0.38}$BN coating, but at somehow higher temperatures. In the as-deposited condition, the peak corresponding to the (200) preferred growth orientation of the cubic nitride phase appeared at even lower diffraction angles than the peak corresponding to stress-free pure $\delta$-TiN, as a result of the residual compressive stresses. Upon annealing at \SI{800}{\celsius}, the peak shifted to higher diffraction angles, due to the relief of compressive residual stresses, and no further changes were observed upon annealing up to \SI{900}{\celsius}. However, at temperatures higher than \SI{1000}{\celsius}, the peak shifted back and tended to asymmetrically broaden towards lower diffraction angles, similarly to what was observed for the Al$_{0.59}$Ti$_{0.41}$N coating at a lower temperature of \SI{850}{\celsius}. This, together with the appearance of small peaks close to the theoretical positions for the (100) and (002) reflections of hexagonal AlN suggests that, in the case of the B doped Al$_{0.62}$Ti$_{0.38}$BN coating, the hexagonal AlN phase emerges at much higher temperatures, \SI{1000}{\celsius}, than for the Al$_{0.59}$Ti$_{0.41}$N coating with a similar $x$ ratio (\SI{850}{\celsius}).

\begin{figure}
	\centering
		\includegraphics[width=\textwidth]{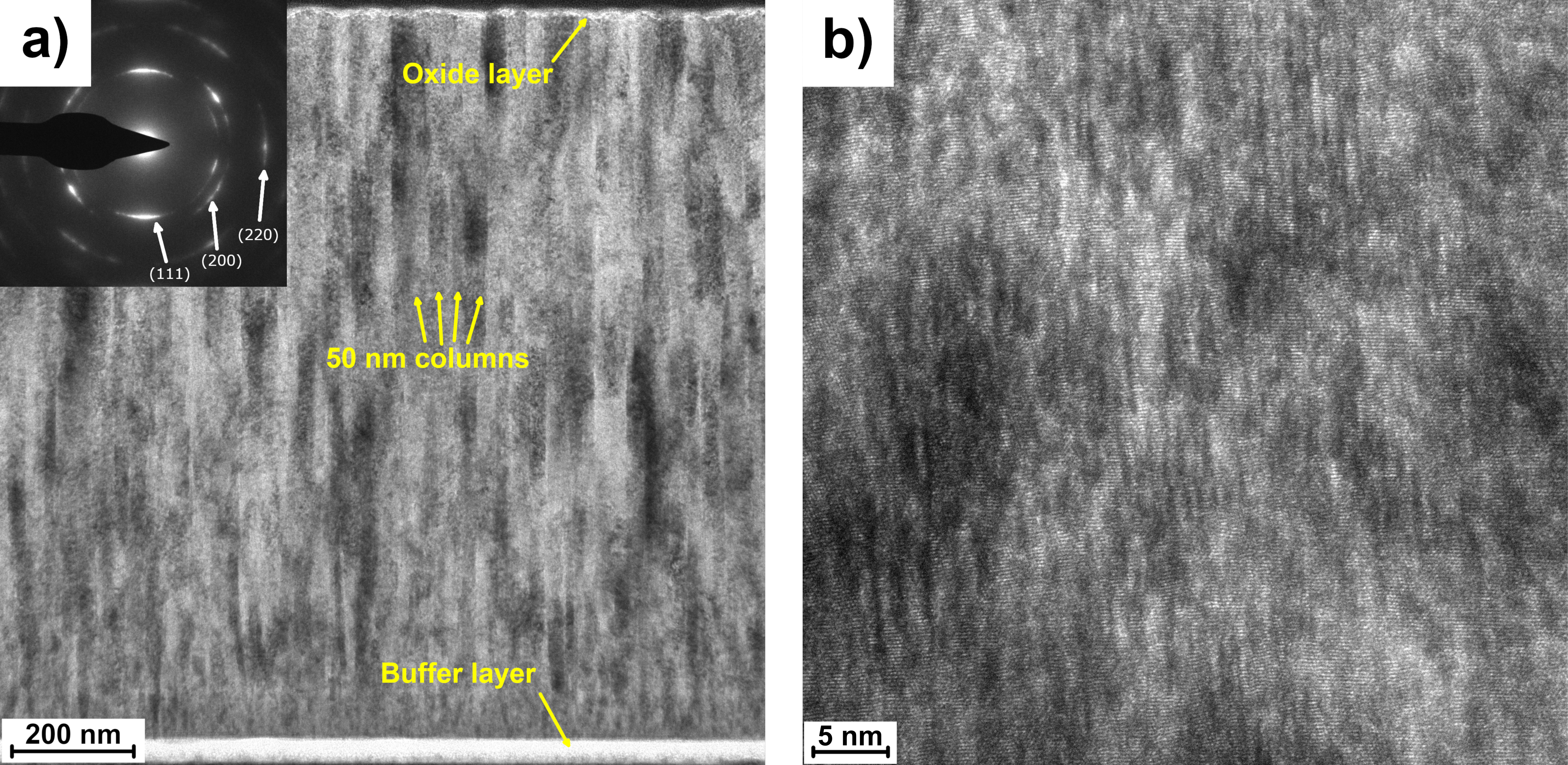} 
		\caption{(a) BF-STEM and (b) HREM image of the Al$_{0.59}$Ti$_{0.41}$N coating after annealing at \SI{900}{\celsius}. The insert in (a) corresponds to the DP obtained from the same area.}
	\label{fig:TiAlN_900C}
\end{figure}

Fig. \ref{fig:TiAlN_900C}(a) shows TEM pictures of the Al$_{0.59}$Ti$_{0.41}$N coating after annealing at \SI{900}{\celsius}. A columnar grain microstructure, similar to that found for the as-deposited condition in Fig. \ref{fig:TiAlN_Combined}, can be seen. The main differences are the formation of a \SI{\approx5}{\nano\meter} thick Al$_2$O$_3$ surface layer, similarly to what has been reported before by McIntyre \textit{et al.} \cite{mcintyre1990}, a slight increase in surface roughness and the width of the columnar grains, up to \SI{\approx50}{\nano\meter}, and a transition in the DP from diffraction spots to discontinuous diffraction rings. The (111) diffraction rings are more intense along the vertical growth direction, in agreement with the preferred growth orientation, but the orientation spread is substantial. This is clearly shown in the HREM of Fig. \ref{fig:TiAlN_900C}(b). The coating remains highly crystalline, as can be seen in the high magnification insert of a selected small region highlighted in yellow. The (111) lattice fringes, visible across the entire area of the image, show some degree of misorientation with respect to the vertical growth direction, in agreement with the angular spread of the diffraction spots in the DP, which might be the result of the relief of compressive residual stresses at high temperature. Finally, and despite what was observed by XRD, no signs of any phase separation or the formation of an hexagonal nitride phase were found by high resolution STEM and/or EELS. However, this might be due to the limitations of the TEM technique, because the domains of phase separation are much smaller than the thickness of the TEM foil.

\begin{figure}
	\centering
		\includegraphics[width=\textwidth]{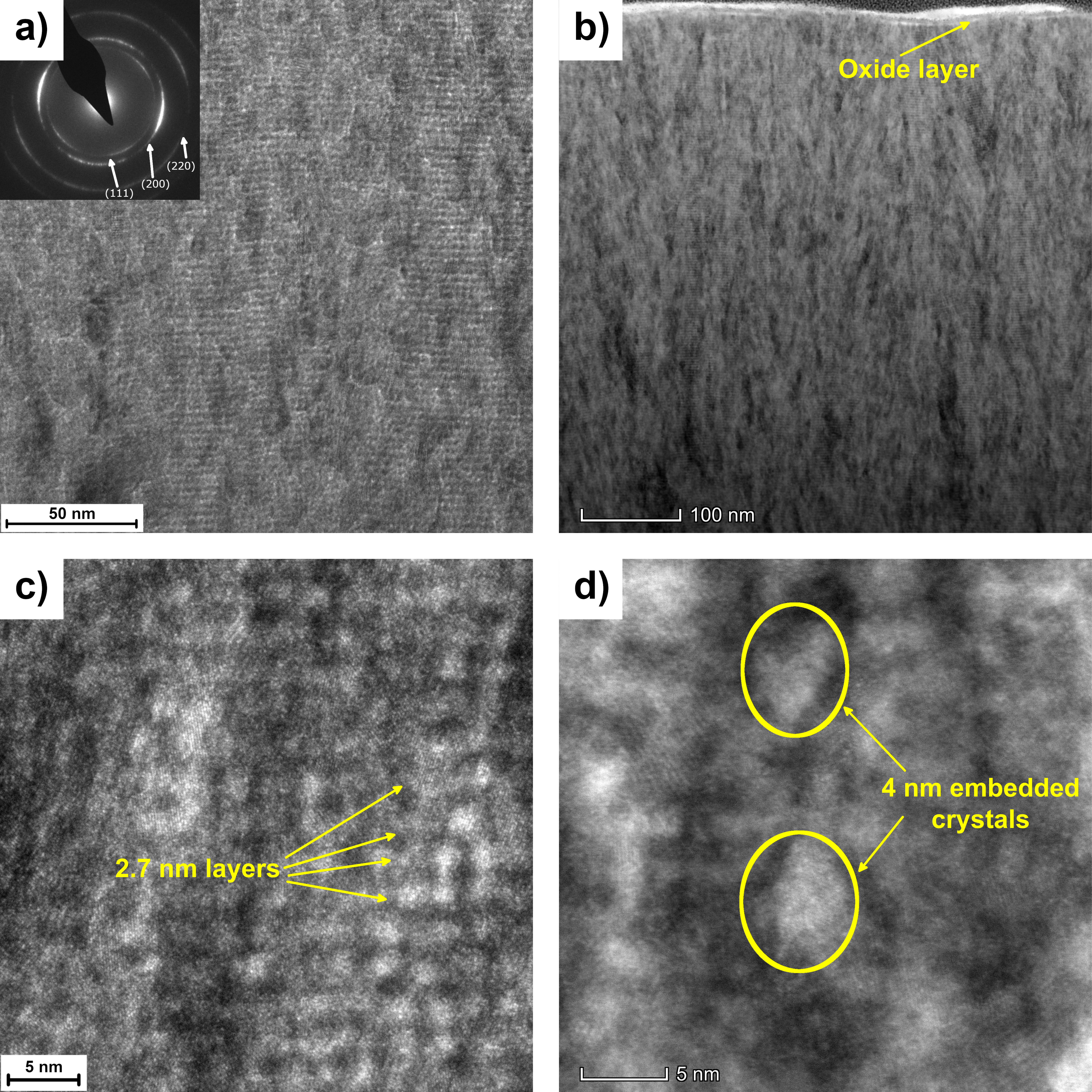} 
		\caption{(a) BF-TEM, (b) BF-STEM, (c) HREM and (d) HAADF-STEM pictures of Al$_{0.62}$Ti$_{0.38}$BN sample after annealing at \SI{900}{\celsius}. The insert in (a) corresponds to the DP obtained from the same area. Growth direction is vertical in all cases.}
	\label{fig:TiAlBN_900C}
\end{figure}

Fig. \ref{fig:TiAlBN_900C}(a) shows a cross-sectional STEM image of the Al$_{0.62}$Ti$_{0.38}$BN coating after annealing at \SI{900}{\celsius}. The microstructure is very similar to that found in the as-deposited coatings, except for the formation of a \SI{\approx5}{\nano\meter} thick Al$_2$O$_3$ surface layer (see Fig. \ref{fig:TiAlBN_900C}(b)). The coating still preserves the multilayered structure, even after annealing at \SI{900}{\celsius} and the microstructure was composed of fine nanocrystalline domains, of size around \SI{4}{\nano\meter}, as evidenced by the diffraction rings seen in the DP shown as an insert in Fig. \ref{fig:TiAlBN_900C}(a), and as highlighted in the HREM and in the HAADF-STEM images (see Figs. \ref{fig:TiAlBN_900C}(c) and (d), respectly).

\begin{figure}
	\centering
		\includegraphics[width=0.7\textwidth]{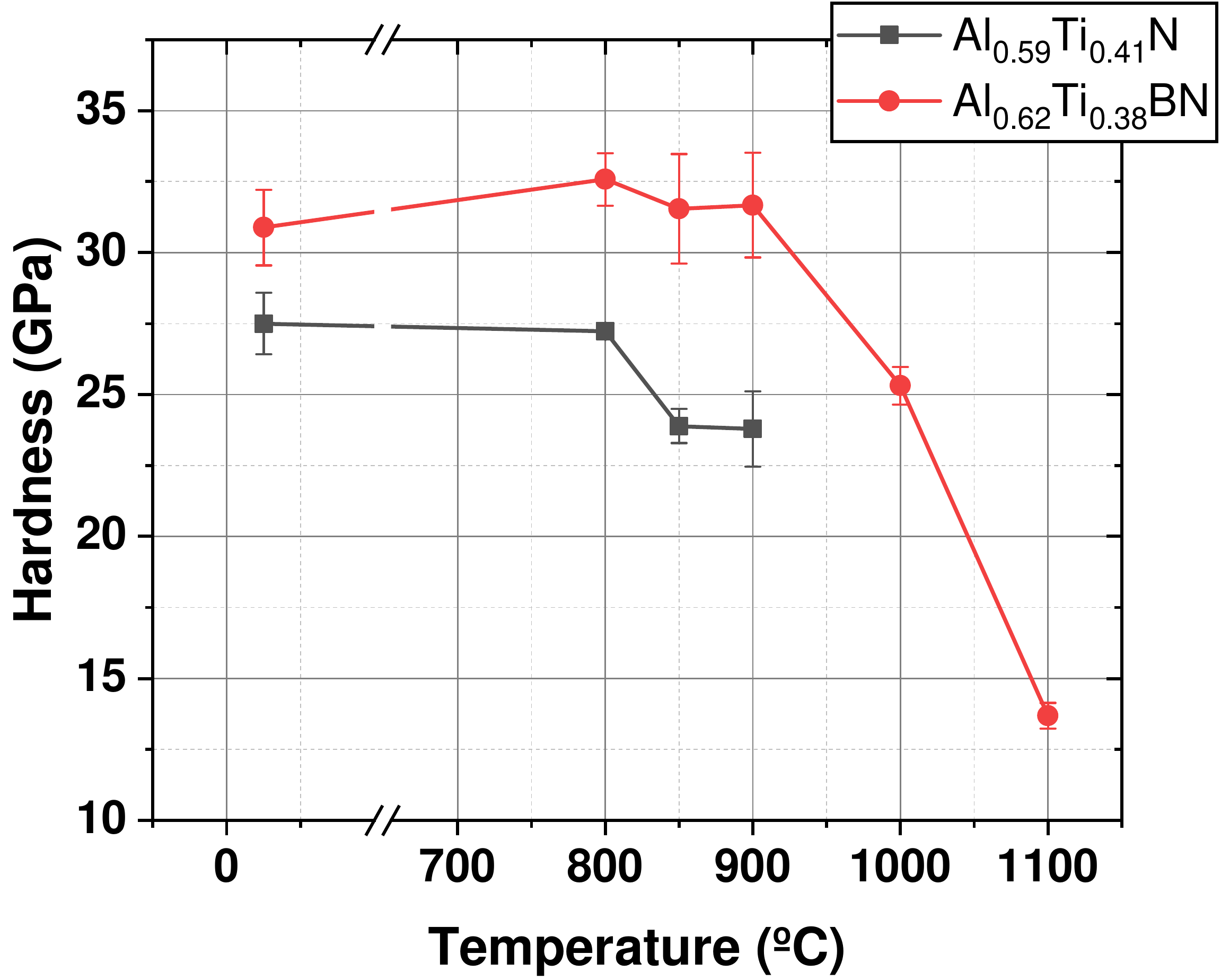} 
		\caption{Evolution of hardness after annealing for Al$_{0.59}$Ti$_{0.41}$N and Al$_{0.62}$Ti$_{0.38}$BN. Al$_{0.62}$Ti$_{0.38}$BN shows improved temperature resistance over Al$_{0.59}$Ti$_{0.41}$N.}
	\label{fig:Annealing_H}
\end{figure}

The evolution of room temperature hardness with annealing temperature is shown in Fig. \ref{fig:Annealing_H}. The Al$_{0.59}$Ti$_{0.41}$N coating maintained a constant hardness even after annealing at \SI{800}{\celsius}, but hardness rapidly decreased for higher temperatures, up to \SI{900}{\celsius}, temperature that produced the delamination of the coating. The corresponding B containing Al$_{0.62}$Ti$_{0.38}$BN coating, in contrast, showed a superior thermal stability. Firstly, it withstood annealing temperatures up to \SI{1100}{\celsius} without delamination. Secondly, it maintained a higher hardness than the Al$_{0.59}$Ti$_{0.41}$N coating for all annealing temperatures. And finally it also showed a slightly increasing hardness in the temperature range between \SI{800}{\celsius} and \SI{1000}{\celsius}, a phenomenon that has been referred to as thermal aging and that has been related to spinodal decomposition of AlTiN based coatings into cubic AlN- and TiN-rich domains \cite{bartosik2017}.

\begin{figure}
	\centering
		\includegraphics[width=0.8\textwidth]{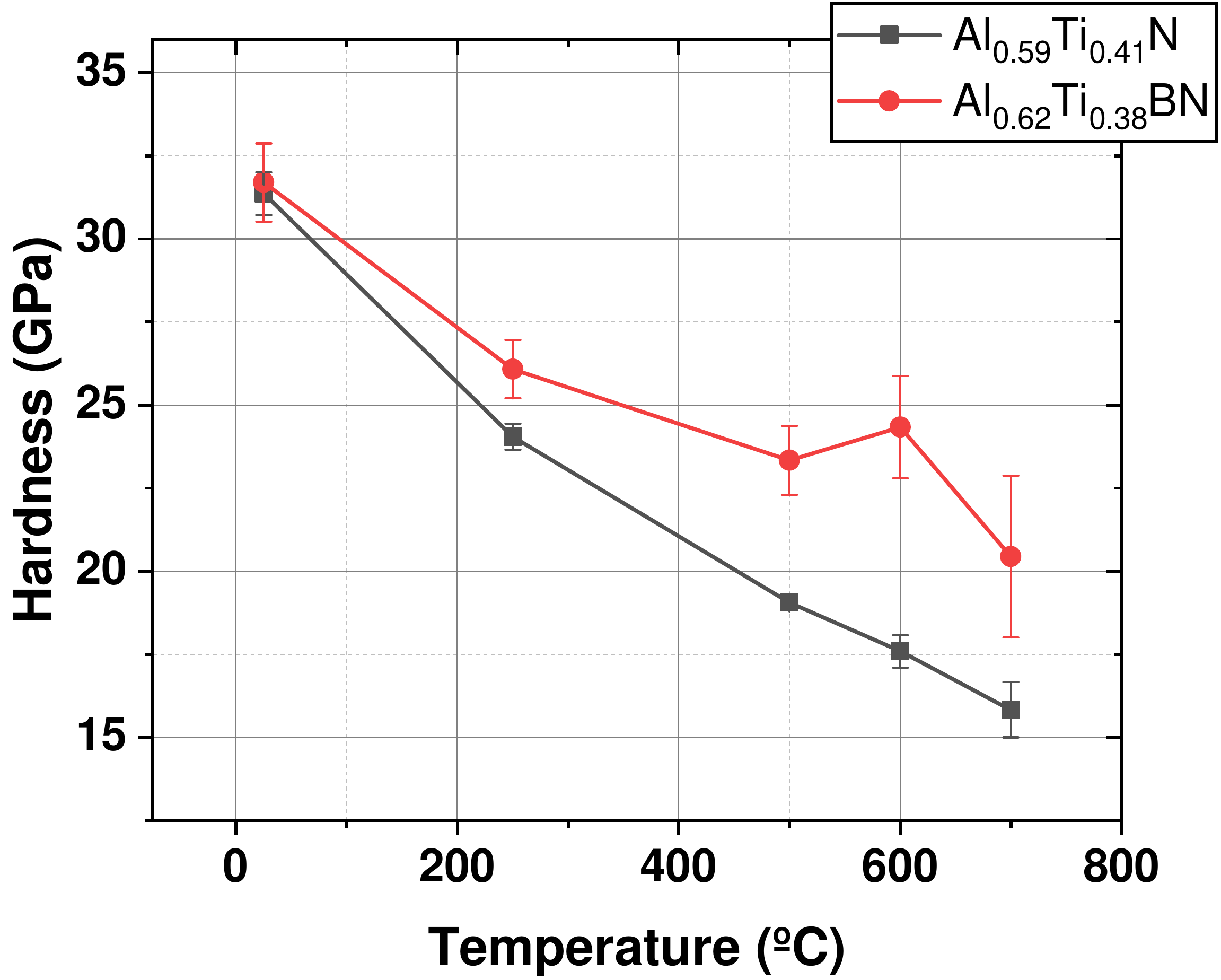} 
		\caption{Hot hardness measurements by nanoindentation for Al$_{0.59}$Ti$_{0.41}$N and Al$_{0.62}$Ti$_{0.38}$BN.}
	\label{fig:HT_Nano}
\end{figure}

The hardness after annealing provides an indication of the thermal stability of the coating, but in order to assess the coating performance under service, hot hardness is a more relevant property. Therefore, the Al$_{0.59}$Ti$_{0.41}$N and Al$_{0.62}$Ti$_{0.38}$BN coatings were also tested by high temperature nanoindentation up to \SI{700}{\celsius} to assess their high temperature mechanical properties. The results are shown in Fig. \ref{fig:HT_Nano}. Both coatings follow a similar trend, experimenting a steady drop in hardness with temperature, due to the thermally activated nature of plastic deformation. However, the hardness decay for the B containing coating was less pronounced, retaining a hardness of \SI{\approx20}{\giga\pascal} at \SI{700}{\celsius}. 

\section{Discussion}\label{sec:Discussion}

As explained in Sec. \ref{sec:Experimental}, the total deposition time and the power applied to the pre-alloyed target used in each coating series was maintained constant in all cases, while the $x$ ratio was adjusted by co-deposition from a supporting target of either Al or Ti at different powers. This can be clearly seen on the total thickness of each coating reported in Table \ref{tab:Composition}, which increases with increasing the power of the supporting target in each case. The $x$ ratio of the AlTiN-2 coating, deposited using only the pre-alloyed Al$_{50}$Ti$_{50}$ target was 0.54, \textit{i.e.}, slightly richer in Al than the target, due to the higher sputtering yield of Al. The same happens for AlTiBN-5, which results in an $x$ ratio of 0.69, slightly richer in Al than the pre-alloyed Al$_{55}$Ti$_{35}$B$_{10}$ target, from which it was deposited. The B content of the AlTiBN-3 coating was estimated by EELS to be \SI{14}{\atpercent}, which is reasonable considering the composition of the Al$_{55}$Ti$_{35}$B$_{10}$ target. Although the B content could not be determined for the rest of the coatings in the series, it is expected that it might be even bigger for the AlTiBN-4 coating deposited without any supporting Ti or Al target. Finally, the reference AlTiBN-DC coating in Table \ref{tab:Composition} has almost twice the thickness of the equivalent coating deposited by HiPIMS (AlTiBN-3) using the same target powers and deposition times, which indicates that the much higher energetic conditions provided by HiPIMS result in much more compact coatings.

The XRD, TEM and nanoindentation results presented above show that B doping has a profound effect on the microstructure and mechanical properties of Al$_x$Ti$_{1-x}$N coatings grown by HiPIMS. In the case of the Al$_x$Ti$_{1-x}$N series, the microstructure and phase composition follow the trends expected from literature. The coatings are composed of a single NaCl-type cubic phase up to Al ratios of $x \approx 0.60$, while higher Al ratios lead to the appearance of some wurtzite-type hexagonal phase, which becomes prominent for $x \approx 0.78$ (see Fig. \ref{fig:TiAlN_TiAlBN_BB}). The stability limit of the metastable Al$_x$Ti$_{1-x}$N cubic phase has been shown to depend on the deposition method. While reactive DC sputter deposition typically produces single-phase cubic Al$_x$Ti$_{1-x}$N coatings for $x \leq 0.5$ \cite{wahlstrom1993, greczynski2014}, the stability limit has been shown to increase up to $x \leq 0.6$, for coatings deposited using methods with a higher degree of plasma ionization, such as cathodic arc evaporation \cite{ikeda1991} or HiPIMS  \cite{greczynski2014, greczynski2012}. In the single-phase stability region, for $x < 0.6$, the coatings maintained a dense columnar grain structure (see Fig. \ref{fig:SEM_Comparison}(a)), with a (111) preferred growth orientation (see Fig. \ref{fig:TiAlN_TiAlBN_BB}). Texture development in Ti(Al)N coatings is affected by many process parameters. It is generally accepted that the (111) preferred orientation arises through a competitive columnar growth process under kinematically limited conditions, because the (111) oriented columnar grains, which are the grains with the fastest growing direction, overlap the rest of the orientations \cite{hultman1995, abadias2008}. Finally, the hardness and elastic modulus peak at $x = 0.6$ to quickly drop for higher Al contents as a result of the hexagonal phase formation (see Fig. \ref{fig:MechProps}), as expected from previous works \cite{greczynski2014}.

In contrast, the Al$_x$Ti$_{1-x}$BN series deposited under identical conditions show a radically different microstructure. The phase composition follows a similar trend and the coatings are composed of a single NaCl-type crystalline cubic phase up to Al ratios of $x \approx 0.60$, which disappears for $x = 0.69$ (Fig. \ref{fig:TiAlN_TiAlBN_BB}(b)), indicating that B addition does not seem to increase the stability limit of the metastable Al$_x$Ti$_{1-x}$N cubic phase. However, for $x \approx 0.60$, B doping results in a dense microstructure formed by nanocrystalline Ti(Al)N domains, with a preferred (200) growth orientation (Fig. \ref{fig:TiAlN_TiAlBN_BB}), and amorphous regions composed of Ti(Al)B$_2$ and BN (Fig. \ref{fig:TiAlBN_Combined} and \ref{fig:EELS}). The size of the crystalline and amorphous domains are of the order of \SI{2}{\nano\meter}, with a phase composition for $x \approx 0.60$ of 77\% Ti(Al)N , 15.7\% BN and 7.3\% Ti(Al)B$_2$. Therefore, B doping results in the suppression of columnar growth (Fig. \ref{fig:SEM_Comparison}(b)) presumably because the BN and Ti(Al)B$_2$ amorphous domains interrupt the growth of the Ti(Al)N crystalline phase, forcing its continuous re-nucleation. Additionally, the deposition strategy used to tailor the Al ratio by co-sputtering from two targets results in a multilayered structure (Fig. \ref{fig:TiAlBN_Combined}), as the rotation of the substrate holder brings the coating surface to alternatively face each of the targets. However, the same strategy was followed in the Al$_x$Ti$_{1−x}$N coating series but, in that case, the growth of the columnar grains was not interrupted as the coating surface faced each target. The latter indicates that the suppression of columnar growth in the case of the Al$_x$Ti$_{1−x}$BN coatings is not a consequence of the co-sputtering from two targets, but the result of the formation of the BN and Ti(Al)B$_2$ amorphous domains. Similar grain refinement effects have also been reported in TiN coatings with small B additions \cite{rother1997}. Additionally, the continuous re-nucleation of the Ti(Al)N cubic-phase is consistent with the observed (200) preferred growth orientation in this series since this is the preferred nucleation orientation of the cubic nitride phase \cite{hultman1995, hultman1989}. Therefore, small additions of B seem to induce a similar effect to that found with Si in TiAlSiN coatings, in which Si contributes to the formation of a nanocomposite structure composed of Ti(Al)N nanocrystalline grains embedded in an amorphous SiN$_x$ matrix \cite{veprek2004}.

With respect to the mechanical properties, the hardness and elastic modulus of both the Al$_x$Ti$_{1−x}$N and Al$_x$Ti$_{1−x}$BN coatings peak for $x \approx 0.6$, as higher Al contents result on the formation of the hexagonal AlN phase (Fig. \ref{fig:MechProps}). However, the increased hardness and reduced modulus of Al$_x$Ti$_{1−x}$BN with respect to the Al$_x$Ti$_{1−x}$N coatings result in larger $H/E$ and $H^3/E^2$ ratios (Fig. \ref{fig:HE_H3E2}) for the former. A higher $H/E$ ratio indicates an improved tribological performance \cite{bull2006}. Moreover, the $H^3/E^2$ ratio gives a rough estimate of whether or not a material will be fracture resistant. The higher $H^3/E^2$ of Al$_x$Ti$_{1−x}$BN, up to $x = 0.62$, suggests that these coatings should exhibit a higher fracture toughness than Al$_x$Ti$_{1−x}$N coatings, which have already been shown to be superior than TiN \cite{bartosik2017}. Additionally, comparison between the Al$_{0.62}$Ti$_{0.38}$BN coating grown by HiPIMS and the Al$_{0.61}$Ti$_{0.39}$BN coating grown by DC magnetron sputtering illustrate that, not only the B content, but also the deposition conditions, and in particular the degree of plasma ionization, play a major role on the suppression of columnar growth and the development of the nanocomposite structure. As a matter of fact, the Al$_{0.61}$Ti$_{0.39}$BN coating grown by DC magnetron sputtering exhibited a (111) texture (Fig. \ref{fig:TiAlN_TiAlBN_BB}(c)) with a porous columnar microstructure and a very low hardness and elastic modulus (Fig. \ref{fig:MechProps}). It is well known that tailoring the ion irradiation conditions (both the flux and energy of bombarding ions) during deposition have profound effects on the microstructure and texture of nitride coatings \cite{adibi1993}. This is even more important in the case of HiPIMS for which a substantial ionization of the sputtered metal flux (Ti$^+$ and $Al^+$) is expected, including the production of multiply-charged metal ions (Ti$^{2+}$ or Al$^{2+}$), which has been shown to be more significant for Ti than for Al \cite{greczynski2012, viloan2020}. This means that, in the coatings under investigation, for which a substrate bias of \SI{−100}{\volt} was used, the energy of the bombarding ions can result in substantial residual lattice damage and undesirable levels of compressive residual stresses. Therefore, there is potentially a large room for improvement on the control of the flux and ion energy of the bombarding ions to optimize the microstructure, texture, residual stresses and mechanical properties of high-Al-content metastable Al$_x$Ti$_{1−x}$BN coatings grown by HiPIMS, although this is beyond the scope of this paper.

Finally, B doping has a profound beneficial effect on the thermal stability and hot hardness of high-Al-content Al$_x$Ti$_{1−x}$N coatings ($x \approx 0.6$). While the Al$_{0.59}$Ti$_{0.41}$N coating remained stable up to \SI{800}{\celsius}, higher temperatures induce the formation of the detrimental hexagonal AlN phase that results in hardness drop and coating delamination at \SI{900}{\celsius}. On the contrary, the formation of the hexagonal AlN phase in the Al$_{0.62}$Ti$_{0.38}$BN coating was delayed to temperatures higher than \SI{1000}{\celsius}. Moreover, annealing in the temperature range between \SI{800}{\celsius} and \SI{900}{\celsius} results in thermal aging, presumably due to spinodal decomposition of the AlTiN phase into cubic AlN- and TiN-rich domains. Finally, Al$_{0.62}$Ti$_{0.38}$BN retains a substantially larger hot hardness (\SI{20}{\giga\pascal}) than Al$_{0.59}$Ti$_{0.41}$N (\SI{15}{\giga\pascal}) at \SI{700}{\celsius}, as a result of the nanocomposite type structure.

\section{Conclusions}\label{Conclusions}

The effect of boron doping on the microstructure, thermal stability and mechanical properties of Al$_x$Ti$_{1-x}$N based coatings deposited by HiPIMS was evaluated for $x$ ratios ranging between 0.5 and 0.7.
The coatings were predominantly formed by a face-centered cubic Ti(Al)N crystalline phase, both with and without B, even for $x$ ratios as high as 0.6. This ratio is higher than that reported for Al$_x$Ti$_{1-x}$N coatings deposited by reactive magnetron sputtering (0.5), which might be a beneficial effect of the highly energetic deposition conditions offered by HiPIMS. 
B doping, in combination with the highly energetic deposition conditions provided by HiPIMS, results in the suppression of the (111) oriented columnar growth typically found in Al$_x$Ti$_{1-x}$N and DC-sputtered coatings. On the contrary, the Al$_x$Ti$_{1-x}$BN coatings grown by HiPIMS present a dense nanocomposite type microstructure, formed by nanocrystalline Ti(Al)N domains, with a preferred (200) growth orientation and amorphous regions composed of Ti(Al)B$_2$ and BN. For high Al contents ($x \approx 0.6$), the B content was \SI{14}{\atpercent}, resulting in a size of the crystalline and amorphous domains of the order of \SI{2}{\nano\meter}, and a phase composition of 77\% Ti(Al)N , 15.7\% BN and 7.3\% Ti(Al)B$_2$. It is hypothesized that the amorphous domains interrupt the growth of the Ti(Al)N crystalline phase, forcing its continuous re-nucleation, and resulting in the nanocomposite type microstructure. 
As a result of B doping, high-Al-content ($x \approx 0.6$) Al$_x$Ti$_{1-x}$BN coatings grown by HiPIMS offer higher hardness, elasticity and fracture toughness than Al$_x$Ti$_{1-x}$N and DC-sputtered coatings. Moreover, the thermal stability is substantially enhanced, delaying the onset of formation of the detrimental hexagonal AlN phase from \SI{850}{\celsius} in the case of Al$_{0.6}$Ti$_{0.4}$N, to \SI{1000}{\celsius} in the case of Al$_{0.6}$Ti$_{0.4}$BN. Finally, B addition also contributes to a superior hot hardness of Al$_{0.6}$Ti$_{0.4}$BN with respect to Al$_{0.6}$Ti$_{0.4}$N coatings, presumably as an additional beneficial effect of the nanocomposite type structure.

\section*{Acknowledgments}\label{Acknowledgements}

This investigation was supported by the Comunidad de Madrid under the Industrial Doctorate program (IND2018/IND-9668).



\begin{sloppypar}
\bibliographystyle{elsarticle-num} 
\bibliography{elsarticle_AlvaroMendez_AlTiBNAnnealing}
\end{sloppypar}




\end{document}